\documentclass{article}

\usepackage{amsfonts}
\usepackage{amsmath}
\usepackage{amssymb}
\usepackage[mathscr]{eucal}
\usepackage{graphicx}

\begin{document}
\title{
\normalsize \hfill UWThPh-2011-27 \\
\normalsize \hfill August 2011 \\ [20mm]
\LARGE Helical motion of elastic spheres
}
\author{
S.~Broda\thanks{Supported by Fonds zur F\"orderung der wissenschaftlichen Forschung in \"Osterreich, Projekt Nr. P20414-N16.}
\\*[3mm]
\small University of Vienna, Faculty of Physics, Gravitational Physics \\
\small Boltzmanngasse 5, 1090 Vienna, Austria
}
\maketitle
\begin{abstract}
We study the elastic deformations that appear due to tidal and centrifugal forces acting on an elastic sphere in helical motion in a spherically symmetric gravitational field, where gravity is considered to be given by either a Newtonian or a Schwarzschild background. We review an existence/uniqueness theorem based on the implicit function theorem for the nonrelativistic case and give explicit solutions to the linearized elastostatic equations in both cases.
\end{abstract}
%
\section{Introduction}
%
In this paper, we consider a sphere composed of homogeneous and isotropic elastic material in helical motion around a fixed gravitational centre (an elastic planet, so to speak, which, like the moon, turns the same side to the gravitational center). Gravity is considered as a given background, described either by a Newtonian potential or the Schwarzschild solution of general relativity. The elastic sphere is therefore exposed to a combination of tidal and centrifugal forces, both of which cause elastic deformations. Due to the helical nature of the motion it is possible to go to the co-rotating coordinate system, in which we face only an elastostatic problem.

The organisation of this paper is as follows: in the sections 2-4 we briefly review the general theory of relativistic elastodynamics, 
relativistic elastostatics and nonrelativistic elastostatics, viewed under one general framework. To get reasonable boundary conditions, in 
elastostatics it turns out it is preferable to use the material picture. In section 5, we consider the general principles 
of existence/uniqueness proofs based on the implicit function theorem in elastostatics, as used by Beig and Schmidt in 
\cite{beig-schmidt-1}, \cite{beig-schmidt-2}, \cite{beig-schmidt-3}. 
After having considered the linearization of the
elastostatic equations in section 5, we review, in section 6, the existence proof given in \cite{beig-schmidt-2} for the elastic sphere on a
circular orbit in a Newtonian gravitational field in a version adapted to our purposes. In section 7, we then give the explicit solution for the linearized
elastostatic equations in this situation, making use of an ansatz from Papkovich and Neuber (\cite{papkovich}, \cite{neuber}, 
also see \cite{lurje} and the references therein). In section 8, we also consider the linearized elastostatic equations for 
the relativistic case and give the  solution. A existence/uniqueness proof similar to the nonrelativistic one is presumably possible, 
but some details still need to be worked out yet. This will be part of a future publication.
%
\section{Relativistic Elastodynamics}
%
We start from full relativistic elastodynamics, as laid down in \cite{beig-schmidt}, \cite{wernig-pichler} or \cite{broda}. The fields of the theory are mappings
\begin{equation}
f : (\mathcal{M}, g_{\mu \nu}) \to (\mathcal{B}, V_{ABC})
\end{equation}
from the spacetime $\mathcal{M}$, equipped with the spacetime metric $g_{\mu \nu}$ (assumed to have signature $(-,+,+,+)$), to the \emph{body manifold} $\mathcal{B}$. It can be thought of as the abstract collection of points ("atoms") the elastic material in question is composed of. Generally, $\mathcal{B}$ is a open, connected and bounded subset of $\mathbb{R}^3$ with smooth boundary $\partial \mathcal{B}$, and equipped with a volume form $V_{ABC}$. Its single componend will be denoted $V(X) = V_{123}(X) > 0$. We will use coordinates $X^A$ on the body manifold $\mathcal{B}$ and $x^\mu$ on the spacetime $\mathcal{M}$.

The mapping $f^A(x^\mu)$, which is also called \emph{configuration}, is required to uniquely define a future-pointing, timelike, normalized ($u^\mu u^\nu g_{\mu \nu} = -1$) 4-velocity vector $u^\mu$ defined by
\begin{equation}
f^A{}_{,\mu} u^\mu = 0
\end{equation}
i.e. the inverse of a point $X^A \in \mathcal{B}$ is a timelike curve in $\mathcal{M}$, the trajectory of that point. We further require $f^A(x^\mu)$ to be orientation-preserving, which allows us to define a positive particle number density $n$ by
\begin{equation}
f^A{}_{,\mu}(x^\sigma) f^B{}_{,\nu}(x^\sigma) f^C{}_{,\rho}(x^\sigma) V_{ABC}(f^D(x^\sigma)) = n \epsilon_{\mu \nu \lambda \rho}(x^\sigma) u^\rho
\end{equation}

Using the configuration mappings $f^A(x^\mu)$ from $\mathcal{M}$ to $\mathcal{B}$ to describe elastic materials is called \emph{spatial} or \emph{Eulerian picture}. This is the more natural way to go in relativistic elastodynamics. It is also possible to use mappings $\phi^i(X^A): \mathcal{B} \to \mathcal{M}$, called \emph{deformations}, which are in a certain sense inverse to the $f^A$s. This is called \emph{material} or \emph{Lagrangian picture}, which is more commonly used in non-relativistic elasticity theory, but in full relativistic elasticity theory, the spatial picture is much more natural. We will later on introduce and use the material picture when dealing with elastostatic situations, though.

From the configuration gradient $f^A{}_{,\mu}$, we can now define the \emph{strain tensor}
\begin{equation}
H^{AB}(x^\lambda) = f^A{}_{,\mu}(x^\lambda) f^B{}_{,\nu}(x^\lambda) g^{\mu \nu} (x^\lambda)
\end{equation}
By definition, it is symmetric and positive definite, so it also has an inverse $H_{AB}$.

An elastic material is specified by giving an \emph{stored-energy function}
\begin{equation}
w = w(f^A, f^A{}_{,\mu}, x^\mu)
\end{equation}
which describes the elastic potential energy per particle. We assume $w$ to be covariant under spatial diffeomorphisms (the so-called ''principle of material frame indifference``), which implies that $w$ depends on the configuration gradient $f^A{}_{,\mu}$ and $x^\mu$ just via the strain tensor (see e.g. \cite{beig-schmidt}):
\begin{equation}
w = w(H^{AB}, f^C)
\end{equation}
From differentiating $w$, we can define the \emph{second Piola-Kirchhoff stress tensor}
\begin{equation}
\sigma_{AB} = - 2 \frac{\partial w}{\partial H^{AB}}
\end{equation}
the \emph{first Piola-Kirchhoff stress tensor}
\begin{equation}
\label{first_piola_kirchhoff_stress_tensor}
\sigma_\mu{}^A = f^B{}_{,\mu} \sigma_{B C} H^{CA}
\end{equation}
and the \emph{Cauchy stress tensor}
\begin{equation}
\sigma_{\mu \nu} = n \  f^A{}_{,\mu} f^B{}_{,\nu} \sigma_{A B} = -2 n \frac{\partial w}{\partial g^{\mu \nu}}
\end{equation}
The equations of motion for the fields $f^A$ are then given as the Euler-Lagrange equations of the action
\begin{equation}
\label{full_elastic_lagrangian}
S = \int_\mathcal{M} n \, \epsilon(H^{AB},f^C) \sqrt{-\det g_{\mu \nu}} \, d^4 x
\end{equation}
where $\epsilon$ is the relativistic total energy density given by rest energy plus stored energy function:
\begin{equation}
\label{relativistic_energy_density}
\epsilon(H^{AB},f^C) = c^2 + w(H^{AB},f^C)
\end{equation}
%

%
\section{Relativistic Elastostatics}
%
\subsection{Material picture}
%
We now want to consider time-independant situations in a way similar to \cite{beig-schmidt-1}. For this, we suppose that we have an everywhere timelike Killing vector field $\xi^\mu$, and that the velocity vector field $u^\mu$ is proportional to it, so that we have
\begin{equation}
\mathcal{L}_{\xi} f^{A} = \xi^{\mu} f^A{}_{,\mu} = 0
\end{equation}
We further suppose that the spacetime $\mathcal{M}$ can be written as product of space and time
\begin{equation}
\mathcal{M} = \mathbb{R} \times N
\end{equation}
where the space manifold $N$ is the quotient of $\mathcal{M}$ by the isometry group generated by $\xi^\mu$. The metric $h_{ij}$ on $N$ is in a suitable coordinate system $(t, x^i)$ on $\mathbb{R} \times N$ given by
\begin{equation}
h_{i j} = g_{i j} - g_{0 i} g_{0 j} / g_{0 0}
\end{equation}
whereas the inverse metric $h^{ij}$ is just equal to the spatial part of the inverse metric of $\mathcal{M}$:
\begin{equation}
h^{i j} = g^{i j}
\end{equation}
This is why we immediately get for the strain tensor
\begin{equation}
\label{static_strain_tensor}
H^{AB} = f^A{}_{,\mu} f^B{}_{,\nu} g^{\mu \nu} = f^A{}_{,i} f^B{}_{,j} h^{i j}
\end{equation}
The particle number density $n$ is now determined via the metric epsilon tensor $\epsilon_{i j k}$ on $N$ belonging to $h_{i j}$ via
\begin{equation}
f^A{}_{,i}(x^l) f^B{}_{,j}(x^l) f^C{}_{,k}(x^l) V_{ABC}(f^D(x^l)) = n \epsilon_{ijk}(x^l)
\end{equation}
which is equivalent to
\begin{equation}
n^2 = \det_{V_{ABC}} ( H^{DE} ) = \frac{V^2 \left( \det{f^A{}_{,k}} \right)^2}{\det h_{i j}}
\end{equation}

To adapt the action \eqref{full_elastic_lagrangian} to the elastostatic setting, we need to split $\sqrt{-\det g_{\mu \nu}}$. To do so,  we use the cofactor formula for the inverse matrix, applied to calculate $g_{00}$ from $g^{\mu \nu}$, which gives
\begin{equation}
g_{0 0} = \frac{1}{\det g^{\mu \nu}} \det h^{i j}
\end{equation}
so the action principle \eqref{full_elastic_lagrangian} can be rewritten as
\begin{equation}
\label{almost_static_action}
S = \int_{\mathbb{R} \times N} n \, \epsilon(H^{AB},f^C) \sqrt{-g_{0 0}} \sqrt{\det h_{i j}} \, d^3 x \, dt
\end{equation}
we rewrite $g_{00}$ further as
\begin{equation}
-g_{00} = c^2 e^{2 U / c^2}
\end{equation}
we insert this in \eqref{almost_static_action}, divide by $c$ and leave the integration over $t$ away to get the reduced action
\begin{equation}
\label{static_action}
S = \int_{N} n \, \epsilon(H^{AB},f^C) e^{U / c^2} \sqrt{\det h_{i j}} \, d^3 x
\end{equation}
By varying this action principle, we get the Euler-Lagrange equations
\begin{equation}
e^{-\frac{U}{c^2}} D_j \left( e^{\frac{U}{c^2}} \sigma_i{}^j \right) - n \left( 1 + \frac{w}{c^2} \right) D_i U = 0
\end{equation}
where $D_j$ is the metric covariant derivative of $h_{ij}$, together with the boundary conditions of \emph{vanishing surface traction}:
\begin{equation}
\label{boundary_conditions_spatial_picture}
\left. \sigma_i{}^j n_j \right|_{f^{-1}(\partial \mathcal{B})} = 0
\end{equation}
with $n_j$ being a co-normal vector to the boundary. Here an awkward feature of the spatial picture shows up: the inverse of $f^A$ appears in the description of the boundary. This difficulty can be avoided by switching to the material picture, which we are now going to introduce.
%
\subsection{Material picture}
%
In the \emph{material} or \emph{Lagrangian} representation of elasticitiy theory,

\begin{equation}
\phi : N \to \mathcal{B}
\end{equation}
In the elastostatic setting, the $\phi^i$ are just the inverse of $f^A$:
\begin{align}
\phi^i(f^A(x^j)) &= x^i,		&	f^A(\phi^i(X^B)) &= X^A
\end{align}
The map $\phi^i(X^A)$ is called \emph{deformation}. From its gradient, we can define another form of the strain tensor:
\begin{equation}
H_{AB}(X^C) = \phi^i{},_A(X^C) \phi^j{},_B(X^C) h_{ij}(\phi(X^C))
\end{equation}
It is the inverse to $H^{AB}$ and as such, contains the same information.

The \emph{first Piola-Kirchhoff stress tensor} already introduced in equation \eqref{first_piola_kirchhoff_stress_tensor} has a particularly simple form in the material picture:
\begin{equation}
\label{first_piola_kirchhoff_stress_tensor_material}
\sigma_i{}^A = f^B{},_i \sigma_{B C} H^{CA} = \frac{\partial w}{\partial \phi^i{}_A}
\end{equation}
It will also appear in the elastostatic equations and the boundary conditions. To get them, we have to transform the action principle \eqref{static_action} to the form
\begin{equation}
\label{material_action}
S = \int_{\mathcal{B}} \epsilon(H^{AB},X^C) e^{U / c^2} \, V d^3 X
\end{equation}
From the variation of \eqref{material_action}, we get the elastostatic equation in the material picture:
\begin{equation}
\label{elastostatic_equations_in_the_material_picture}
e^{-U/c^2} \nabla_A \left( e^{U/c^2} \sigma_i{}^A \right) - \left( 1 + \frac{w}{c^2} \right) \partial_i U = 0
\end{equation}
where $\nabla_A$ is the so-called \emph{material derivative} given by
\begin{equation}
\nabla_A \sigma_i{}^A = \frac{1}{V} \partial_A \left( V \sigma_i{}^A \right) - \Gamma^k_{i j} \sigma_k{}^A \phi^i{},_A
\end{equation}
The $\Gamma^k_{i j}$ denote the Christoffel symbols of the metric $h_{ij}$, evaluated at $\phi^m(X^B)$.

The elastostatic equations in the material picture \eqref{elastostatic_equations_in_the_material_picture} are accompanied by the following boundary conditions:
\begin{equation}
\label{boundary_conditions_in_the_material_picture}
\left. \sigma_i{}^A n_A \right|_{\partial \mathcal{B}} = 0
\end{equation}
where $n_A$ is a conormal vector to $\partial \mathcal{B}$. As already noted, the structure of this boundary conditions is much nicer than in the spatial picture (equation \eqref{boundary_conditions_spatial_picture}), because the boundary is fixed and does not depend on the fields.

%
\section{Nonrelativistic Elastostatics}
%
The equations governing nonrelativistic elastostatic can be retrieved as limit for $\frac{1}{c} \to 0$, which can be applied directly to the action principle \eqref{material_action} using the decomposition \eqref{relativistic_energy_density} and
\begin{equation}
e^{U/c^2} = 1 + \frac{U}{c^2} + \mathcal{O} \left( \frac{1}{c^4} \right)
\end{equation}
if one neglects a constant ''Null Lagrangian`` proportional to $c^2$, which goes to infinity in the nonrelatvistic limit, but doesn't contribute to the resulting Euler-Lagrange equations to get
\begin{equation}
\label{nonrelatvistic_action}
S = \int_{\mathcal{B}} \left( w(H^{AB},X^C) + U(\phi^i(X^C))  \right) \, V d^3 X
\end{equation}
From varying this action, we get the nonrelativistic elastostatic equations
\begin{equation}
\label{nonrelativistic_equation}
\nabla_A \sigma_i{}^A - \partial_i U = 0
\end{equation}
again together with the boundary conditions
\begin{equation}
\label{nonrelativist_boundary_condition}
\left. \sigma_i{}^A n_A \right|_{\partial \mathcal{B}} = 0
\end{equation}
Alternatively, those could also have been retrieved by applying the limit $\frac{1}{c} \to 0$ directly to \eqref{elastostatic_equations_in_the_material_picture}, \eqref{boundary_conditions_in_the_material_picture}.
%
\section{Existence/Uniqueness Proof}
%
%
\subsection{Reference State}
%
For the purpose of finding existence/uniqueness results, we assume that there is a stress-free reference state $\bar \phi^i$ (all quantities referring to this reference configuration will be denoted with a bar). We will later use the implicit function theorem for finding solutions near the reference state. Stressfreeness (i.e. $\bar \sigma_i{}^A = 0$) is equivalent to (because of equation \eqref{first_piola_kirchhoff_stress_tensor}):
\begin{equation}
\left. \frac{\partial w(H^{CD},X^E)}{\partial H^{AB}} \right|_{\phi = \bar \phi} = 0
\end{equation}
We can always choose coordinate systems so that the reference deformation map is just the identity map on $\mathcal{B}$, i.e. we identify $\mathcal{B}$ with the subset of space $N$ occupied by the elastic body in the reference state:
\begin{equation}
\bar \phi^i(X^A) = \delta^i_A X^A
\end{equation}
In such a coordinate system, $H^{AB}=\delta^{AB}$ can be considered as metric on $\mathcal{B}$. We further assume that the material has a constant density $\rho_0$ in the reference state, which is equivalent to the volume form on $\mathcal{B}$ being the metric one and its component $V(X)=const$.
%
\subsection{Operator formulation}
\label{operator_formulation}
%
We can write equation \eqref{nonrelativistic_equation} in the form
\begin{equation}
\label{operator_equation}
\mathscr{E}_i[\phi^j] + \mathscr{F}_i[\phi^j] = 0
\end{equation}
where the \emph{elasticity operator} $\mathscr{E}_i[\phi^j]$ is defined by
\begin{equation}
\label{elasticity_operator}
\mathscr{E}_i[\phi^j] = \nabla_A \sigma_i{}^A
\end{equation}
It is quasi-linear, as can be easily calculated from formula \eqref{first_piola_kirchhoff_stress_tensor_material}:
\begin{equation}
\begin{split}
\partial_A \sigma_i{}^A &= \left( 4 H^{AE} H^{BF} f^C{}_{,i} f^D{}_{,j} L_{CEDF} - f^C{}_{,i} H^{AB} H_{CD} \sigma_j{}^D - \right. \\
& \qquad \left. - f^B{}_{,i} \sigma_j{}^A - f^B{}_{,j} \sigma_i{}^A \right) \left( \phi^j{}_{,AB} \right) + l.o.
\end{split}
\end{equation}
where the \emph{elasticity tensor} $L_{ABCD}$ is given by
\begin{equation}
L_{ABCD} = \frac{\partial^2 w}{\partial H^{AB} \partial H^{CD}}
\end{equation}
For its value in the reference state, we require the \emph{uniform pointwise stability condition}
\begin{equation}
\label{stability_condition}
\bar L_{ABCD} M^{AB} M^{CD} > 0   \qquad \forall M^{AB} = M^{(AB)} \not= 0
\end{equation}
In case of homogeneous and isotropic materials, the elasticity tensor can be expressed using the Lam\'e constants $\lambda$ and $\mu$ and the density $\rho_0$ of the material in the reference state:
\begin{equation}
\label{definition_L_ABCD}
\bar L_{ABCD} = \frac{1}{4 \rho_0} \left( \lambda \delta_{AB} \delta_{CD} + \mu \left( \delta_{AC} \delta_{BD} + \delta_{AD} \delta_{BC} \right) \right)
\end{equation}
For the Lam\'e constant, the condition \eqref{stability_condition} is equivalent to
\begin{equation}
3 \lambda + 2 \mu > 0, \qquad \mu > 0
\end{equation}

The force term in equation \eqref{operator_equation} is just given as \emph{Nemitskii} or \emph{composition operator}:
\begin{equation}
\label{force_operator}
\mathscr{F}_i[\phi^j](X^A) = (K_i \circ \phi)(X^A)
\end{equation}
where $K_i = - \partial_i U$, as usual. It is also called ''load`` in the context of elastostatics.

It will turn out to be convenient to extend equation \eqref{operator_equation} in order to contain the boundary conditions as well. We thus introduce operators $E, F: \mathscr{C} \to \mathscr{L}$, which assign tuples $(b_i, \sigma_i) \in \mathscr{L}$ to deformation maps $\phi^i \in \mathscr{C}$, where $b_i$ is a volume force definded on $\mathcal{B}$ and $\sigma_i$ is a surface force defined on $\partial \mathcal{B}$.

In order for the operators $E$ and $F$ to be well-defined and $C^1$, we assume the spaces $\mathscr{C}$ and $\mathscr{L}$ to be the Sobolev spaces standard in elasticity theory (see \cite{valent}, \cite{ciarlet}): for $\mathscr{C}$ we use a neighbourhood of the reference configuration $\bar \phi^i$ in $W^{2,p}(\mathcal{B}, \mathbb{R}^3)$ with $p>3$, which we assume to be small enough so that every $\phi^i \in \mathscr{C}$ has an $C^1$-inverse. For $\mathscr{L}$, which is also called \emph{load space}, we use
\begin{equation}
\mathscr{L} = W^{0,p}(\mathcal{B}, \mathbb{R}^3) \times W^{1-1/p,p}(\partial \mathcal{B}, \mathbb{R}^3)
\end{equation}
Thus, we now have to solve the equation
\begin{equation}
\label{extended_operator_equation}
E[\phi^j] + t F[\phi^j] = 0
\end{equation}
where $E$ and $F$ are given by
\begin{equation}
\label{definition_E}
E[\phi](X^B) = ((\nabla_A \sigma_i{}^A)(X^B), (\sigma_i{}^A n_A)(X^B)|_{\partial \mathcal{B}})
\end{equation}
and
\begin{equation}
\label{definition_F}
F[\phi](X^B) = ((K_i \circ \phi) (X^B), 0)
\end{equation}
the second component is just 0, because we always use the boundary condition of vanishing surface traction (equation \eqref{boundary_conditions_in_the_material_picture}). It is well known that such an composition operator is $C^1$ if the vector field $K_i$ is (see \cite{valent}). The variable $t$ will become our linearization parameter.

An important property of the operator $E$ is that it does not have full range: if a load $(b_i, \sigma_i) \in E(\mathscr{C})$, it has to satisfy the \emph{equilibrium conditions} for the six Euclidean Killing vectors $\xi^i$
\begin{equation}
\label{equilibrium_conditions}
\int_{\mathcal{B}} (\xi \circ \phi)(X) b_i(X) dV(X) + \int_{\partial \mathcal{B}} (\xi \circ \phi)(X) \sigma_i(X) dS(X) = 0
\end{equation}
This is most conveniently seen in the spatial picture by using the symmetry of $\sigma^{i j}$, the Killing equation for $\xi^i$ and Stokes' theorem. The intuitive meaning of this is that the total force and total torque on $\mathcal{B}$ have to vanish, otherwise the body would be set into motion, and no static solutions would exist.

Finally let us review some well-known facts about the linearization $\delta E$ at $\phi^i = \bar \phi^i$ (see \cite{valent}): it is a Fredholm operator with a kernel given by elements of the form
\begin{equation}
\delta \phi^i(X^A) = (\xi^i \circ \bar \phi)(X^A)
\end{equation}
where $\xi^i$ runs through the six Euclidean Killing vectors. Also the image of $\delta E$ has co-dimension six, and is given by the subspace of $\mathscr{L}$ that satisfies the equilibrium conditions \eqref{equilibrium_conditions} for $\phi=\bar \phi$.

We now want to invoke the implicit function theorem to prove existence and uniqueness of solutions of \eqref{extended_operator_equation}. This is possible, because we have already seen that both operators $E$ and $F$ are $C^1$ in the appropriate sense. We are thus considering one-parametric families $\phi^i_t$ of solutions ($t$ being the parameter) with $\phi^i_0 = \bar \phi^i$. As a first step, we note that we obviously have $E[\bar \phi^i] = 0$ because of the stressfreeness of the reference configuration. But unfortunately, since the linearization of the operator $E$ at $\bar \phi^i$ has a non-trivial kernel and range, we cannot apply the implicit function theorem directly. We circumvent this difficulty by a two-step procedure:

The first step is to restrict $\mathscr{C}$ and $\mathscr{L}$ to $\tilde{\mathscr{C}}$ and $\tilde{\mathscr{L}}$ so that the linearization of $E$ becomes an isomorphism. To deal with the non-trivial range of $E$, we define $\tilde{\mathscr{L}} := E(\mathscr{L})$ and choose a fixed 6-dimensional complement $S_6$ of it, i.e.
\begin{equation}
\mathscr{L} = E(\mathscr{L}) \oplus S_6
\end{equation}
This defines a unique projection operator $\mathbb{P}$ from $\mathscr{L}$ to $\tilde{\mathscr{L}}$ along $S_6$:
\begin{equation}
\mathbb{P}: \tilde{\mathscr{L}} \oplus S_6 \to \tilde{\mathscr{L}}
\end{equation}
We now apply $\mathbb{P}$ to equation \eqref{extended_operator_equation}
\begin{equation}\label{projected extended_operator_equation}
\mathbb{P} \circ E[\phi^j] + \mathbb{P} \circ F[\phi^j] = 0
\end{equation}
The linearization of $\left( \mathbb{P} \circ E \right) [\phi^j]$ is just $\left( \mathbb{P} \circ \delta E \right)[\delta \phi^j]$, which is onto $\tilde{\mathscr{L}}$ by definition.

To solve the problem of the non-trivial kernel, we consider solutions of the form
\begin{equation}
\label{phi_kerne_decomposition}
\phi^i_t(X^A) = \xi^i_t + \tilde \phi^i_t(X^A)
\end{equation}
where $\xi^i_t$ is an arbitrary element from the kernel of $E$ and $\tilde \phi^i$ is from a complement of the kernel of $E$ that contains $\bar \phi^i$, which we will denote $\tilde{\mathscr{C}}$.

With this restriction, $\mathbb{P} \circ \delta E$ is now an isomorphism from $\tilde{\mathscr{C}}$ to $\tilde{\mathscr{L}}$ for each $\xi^i_t$ hold fixed, and we can apply the implicit function theorem to obtain a unique solution for $\tilde \phi^i(X^A)$.

As second step, it remains to show that it is possible to find $\xi^i_t$ so that the equilibrium conditions \eqref{equilibrium_conditions} are satisfied. In other words, one needs to check that the full equations \eqref{extended_operator_equation} rather than just the projected 
ones \eqref{projected extended_operator_equation} are valid by a suitable choice of $\xi^i_t$. This can then be done using the finite-dimensional implicit function theorem. For details, we refer to the concrete example below.

%
\section{Linearization}
%
The perturbation of $\phi^i$ is defined as
\begin{equation}
\delta \phi^i = \left. \frac{d \phi^i_t}{d t}   \right|_{t=0}
\end{equation}
From perturbing \eqref{static_strain_tensor} we get the expression
\begin{equation}
\delta H^{AB} = \delta f^A{}_{,i} f^B{}_{,j} h^{i j} + f^A{}_{,i} \delta f^B{}_{,j} h^{i j} = - 2 \delta^A_i \delta^B_j \delta \phi^{(i,j)}
\end{equation}
The perturbation for the first Piola-Kirchhoff stress tensor can be calculated from \eqref{first_piola_kirchhoff_stress_tensor_material}:
\begin{equation}
\label{perturbed_first_piola_kirchhoff_stress_tensor}
\delta \sigma_i{}^A = -2 \bar H^{AB} \bar f^C{}_{,i} \bar L_{BCDE} \delta H^{DE}
\end{equation}
Here, it was used that the reference configuration is stress-free by assumption; otherwise, additional terms would appear in \eqref{perturbed_first_piola_kirchhoff_stress_tensor}, which would make our treatment much more complicated.

Using this, the definition of $\bar L_{ABCD}$ (equation \eqref{definition_L_ABCD}) and our assumption on the coordinate system ($\bar \phi^i{}_{,A} = \delta^i_A)$, we get
\begin{equation}
\delta \sigma_{iA} = \frac{1}{\rho_0} \left( \lambda \delta_{iA} \partial_j \delta \phi^j + \mu \left( \delta \phi_{i,A} + \delta \phi_{A,i} \right) \right)
\end{equation}
which leads to the standard expression of the linearized elasticity operator
\begin{equation}
\partial_A \delta \sigma_i{}^A = \frac{1}{\rho_0} \left( (\lambda + \mu) \partial_i \partial_j \delta \phi^j + \mu \Delta \delta \phi_i \right)
\end{equation}
In order to get correct physical dimensions, we define the \emph{displacement vector} as
\begin{equation}
u^i(x^j) = t \delta \tilde \phi^i(\bar f^A(x^j))
\end{equation}
We have thus arrived at the standard equation of linearized elastostatics:
\begin{equation}
\mu u_{i,jj} + (\lambda + \mu) u_{k,ki} + t \rho_0 K_i = 0
\end{equation}
subject to the boundary conditions
\begin{equation}
\delta \sigma_i{}^A n_A|_{\partial \mathcal{B}} = 0
\end{equation}
%

%
\section{Circular orbits in Newtonian Gravity}
%
We now consider a spherical elastic body with radius $a$ on a circular orbit around a fixed center with mass $M$ in the distance $L$ in the case of Newtonian gravity. The force density is given by $K_i = - \partial_i U$ with
\begin{equation}
U = - \left( \frac{\omega^2 (\mathbf{x}-(\mathbf{x},\mathbf{m}) \mathbf{m})^2}{2} + \frac{GM}{r} \right)
\end{equation}
where $\mathbf{m}$ is a unit vector orthogonal to the orbit plane and $\mathbf{n}$ is the unit vector connecting the body to the force center, and $\omega$ is the angular frequency. Obviously, the force field $K_i$ vanishes on a circle given by $r=L$ if $L$ satisfies the relation
\begin{equation}
\omega^2 L = \frac{G M}{L^2}
\end{equation}
On this orbit, we can write the potential as
\begin{equation}
\label{full_potential}
U = - \omega^2 \left( \frac{(\mathbf{x}-(\mathbf{x},\mathbf{m}) \mathbf{m})^2}{2} + \frac{L^3}{r} \right)
\end{equation}
We will use $\omega^2=\frac{G M}{L^3}$ as our linearization parameter $t$.

The whole threatment of this problem can be simplified by utilizing the mirror symmetry along the planes orthogonal to $\mathbf{n}$ and $\mathbf{n} \times \mathbf{m}$. In a cartesian coordinate system with the axes $(-\mathbf{n}, \mathbf{n} \times \mathbf{m}, \mathbf{m})$, this means:
\begin{equation}
\label{mirror_symmetries}
\begin{split}
\phi^1(X^1, X^2, X^3) = \phi^1(X^1, -X^2, X^3) = \phi^1(X^1, X^2, -X^3) \\
\phi^2(X^1, X^2, X^3) = -\phi^2(X^1, -X^2, X^3) = \phi^2(X^1, X^2, -X^3) \\
\phi^3(X^1, X^2, X^3) = \phi^3(X^1, -X^2, X^3) = -\phi^3(X^1, X^2, -X^3) \\
\end{split}
\end{equation}
We can restrict the spaces $\mathscr{C}$ and $\mathscr{L}$ of configurations and loads to those also satisfying this conditions, and denote them $\mathscr{C}_{sym}$ and $\mathscr{L}_{sym}$). This is possible, because $\mathscr{C}_{sym}$ still contains the identity, which we use as $\bar \phi^i$. Because of the assumption of homogeneity and isotropy, the operators $E$ and $F$ can be restricted to go $\mathscr{C}_{sym} \to \mathscr{L}_{sym}$.

The only Killing vector that satisfies the conditions \eqref{mirror_symmetries} is the translational one $n^i$ (pointing from the body to the gravitational centre). This generates the Kernel of $E:\mathscr{C}_{sym} \to \mathscr{L}_{sym}$, i.e. it is only one-dimensional. Also, the equilibrium conditions \eqref{equilibrium_conditions} are all automatically satisfied except for $\xi^i = n^i$, i.e. also the co-range of $E:\mathscr{C}_{sym} \to \mathscr{L}_{sym}$ only has dimension $1$. We thus have to consider just $\phi^i$ of the following form
\begin{equation}
\label{phi_kerne_decomposition_special_case}
\phi^i_t(X^A) = C(t) n^i + \tilde \phi^i_t(X^A)
\end{equation}
However, the first step of the proof from subsection \ref{operator_formulation} still remains valid in analogous terms. We thus have existence and uniqueness for $\tilde \phi^i_t(X^A)$. As second step, it remains now to determine $C(t)$ from the equilibrium condition
\begin{equation}
\label{equilibrium_condition_z}
N \left( t, C(t) \right) = \int_\mathcal{B} n^i \left( K_i \circ \phi \right) (X) d^3 X
\end{equation}
with $K_i$ being the force field, divided by the linearization constant:
\begin{equation}
K_i = \partial_i \left( \frac{1}{2} \left( \mathbf{x}^2 - \left( \mathbf{x}, \mathbf{m} \right)^2 \right) + \frac{L^3}{r} \right)
\end{equation}
We again use the implicit function theorem; but this time, the finite-dimensional one is sufficient. First we need to show that $N(0,0)=0$, i.e. that the force field $K_i$ is equilibrated at the reference state $\bar \phi^i$. To do so, we use that the integrand in \eqref{equilibrium_condition_z} is a harmonic function
\begin{equation}
\Delta \left( n^i K_i(x) \right) = n^i \Delta K_i(x) = 0
\end{equation}
We further use that $\mathcal{B}$ is a ball; we can thus invoke the mean value theorem for harmonic functions (see e.g. \cite{evans}) to get
\begin{equation}
N(0,0) = \int_\mathcal{B} n^i K_i(x) d^3 x = |\mathcal{B}| \left( n^i K_i \right) (x_0) = 0
\end{equation}
where we have used that the center $x_0$ of $\mathcal{B}$, the force field $K_i$ vanishes.

Now it remains to show that $\frac{\partial N}{\partial C}(0,0) \ne 0$. From \eqref{equilibrium_condition_z} we calculate (using  the decomposition \eqref{phi_kerne_decomposition})
\begin{equation}
\frac{\partial N}{\partial C} \left( 0, 0 \right) = \int_\mathcal{B} \left( \left( K_i n^i \right)_{,j} n^j \right) d^3 X
\end{equation}
Again, the integrand is a harmonic function, so the integral can be evaluated using the mean value theorem for harmonic functions; an easy calculation shows that $\left( \left( K_i n^i \right)_{,j} n^j \right)(x_0) = 3$, so we get
\begin{equation}
\label{N_by_C}
\frac{\partial N}{\partial C} \left( 0, 0 \right) = 3 |\mathcal{B}| = 4 \pi a^3
\end{equation}
which is non-vanishing, as required by the implicit function theorem. We have thus shown existence and uniquenes of both $C(t)$ and $\tilde \phi^i_t(X^A)$, hence $\phi^i_t(X^A)$, for small values of the parameter $t=\omega^2$.
%
\section{Explicit solution}

%
For convenience, we use the linearized strain tensor, which is given by
\begin{equation}
\label{linearized_strain_tensor}
H_{i j} = u_{(i,j)}
\end{equation}
We will use spherical coordinates with normalized coordinate basis vectors; in such a coordinate system, we get for the components of \eqref{linearized_strain_tensor}
\begin{equation}
\begin{split}
H_{r r} &= \frac{\partial u_r}{\partial r} \\
H_{\theta \theta} &= \frac{1}{r} \frac{\partial u_\theta}{\partial \theta} + \frac{u_r}{r} \\
H_{\phi \phi} &= \frac{1}{r} \left( \frac{\cos \theta}{\sin \theta} u_\theta + u_r + \frac{1}{\sin \theta} \frac{\partial u_\phi}{\partial \phi} \right) \\
H_{r \theta} &= \frac{1}{2} \left( \frac{1}{r} \frac{\partial u_r}{\partial \theta} + r \frac{\partial}{\partial r} \left( \frac{u_\theta}{r} \right) \right) \\
H_{r \phi} &= \frac{1}{2} \left( \frac{\partial u_\phi}{\partial r} - \frac{u_\phi}{r} + \frac{1}{r \sin \theta} \frac{\partial u_r}{\partial \phi} \right) \\
H_{\theta \phi} &= \frac{1}{2 r} \left( \frac{\partial u_\phi}{\partial \theta} - \frac{\cos \theta}{\sin \theta} u_\phi + \frac{1}{\sin \theta} \frac{\partial u_\theta}{\partial \phi} \right) \\
\operatorname{tr} H &= \operatorname{div} \mathbf{u} = \frac{1}{r^2} \frac{\partial}{\partial r} \left( r^2 u_r \right) + \frac{1}{r \sin \theta} \frac{\partial}{\partial \theta} \left( u_\theta \sin \theta \right) + \frac{1}{r \sin \theta} \frac{\partial u_\phi}{\partial \phi}
\end{split}
\end{equation}
the linearized stress tensor $\delta \sigma_{i A}$ is then connected to the strain tensor via the generalized Hooke's law for homogeneous and isotropic materials:
\begin{equation}
\delta \sigma_{i A} = 2 \mu H_{i j} \delta^j_A + \lambda \operatorname{tr} H \delta_{i A}
\end{equation}
We now use the following strategy to solve the linearized elastostatic equations with the force given by the potential \eqref{full_potential}: we first decompose the potential in two parts in subsection \ref{force_decomposition_subsection}, where both parts are axisymmetric, and one being equivalent to the problem of a self-rotating sphere, already solved in \cite{beig-schmidt-1}. Because of linearity, both parts can be threated independantly; we concern ourself only with the yet unsolved part of the problem. Using the assumption of axisymmetry, it is relatively easy to find a particular solution first (in subsection \ref{particular_solution_subsection}), that is, a solution to the inhomogeneous linearized elastostatic equations which doesn't necessarily also satisfy the boundary conditions yet. In a second step (subsection \ref{homogeneous_solution_subsection}), we then find, using again axisymmetry and the ansatz from Papkovich and Neuber, the general solution to the homogeneous elastostatic equations in terms of a Legendre polynomial series. ''General`` here means that the homogeneous solution contains enough yet unspecified constants so it may be matched to any boundary conditions. Then the particular solution and the homogeneous solution are added and the constants specified by imposing the zero-traction boundary condition.
%
\subsection{Force decomposition}
\label{force_decomposition_subsection}
%
The potential \eqref{full_potential} has the disadvantage of being not rotationally symmetric, but it can be decomposed in two parts with that property by a simple coordinate transformation: a shift of the coordinate origin to the center of the body at $-L \mathbf{n}$ (i.e. by $\mathbf{y} = \mathbf{x} + L \mathbf{n}$). The centrifugal term in the potential then becomes
\begin{equation}
\begin{split}
- \frac{1}{2} \left( \mathbf{x} - (\mathbf{x}, \mathbf{m}) \mathbf{m} \right)^2 & = - \frac{1}{2} \left( \mathbf{x}^2 - (\mathbf{x}, \mathbf{m})^2 \right) = \\
& = - \frac{1}{2} \left( y^2 - 2 L (\mathbf{y}, \mathbf{n}) + L^2 - (\mathbf{y}, \mathbf{m})^2 \right)
\end{split}
\end{equation}
The term $-\frac{1}{2} \left( y^2-(\mathbf{y},\mathbf{m})^2 \right)$ is just the centrifugal potential caused by a rotation of the body along an axis through its center; this problem has already been solved (see \cite{beig-schmidt-1}, \cite{love}). In the end, this solutions can be just added because of the linearity of the linearized elastic equation. The constant term proportional to $\frac{L^2}{2}$ can be neglected because it doesn't contribute to the force density. The remaining term
\begin{equation}
\label{interesting_potential_term}
L (\mathbf{y}, \mathbf{n}) = L y P_1(\cos \theta)
\end{equation}
is the part of the potential that is going to be used further on. It has the advantage of being harmonic and symmetric under rotations along the $\mathbf{n}$-axis.

Since we have $y<L$, we can expand the Newtonian potential in Legendre polynomials:
\begin{equation}
\label{one_over_r_expansion}
- \frac{1}{|\mathbf{y}-L \mathbf{n}|} = - \frac{1}{L \sqrt{1 - 2 \frac{y}{L} \cos \theta + \frac{y^2}{L^2}}} = - \frac{1}{L} \sum_{n=0}^\infty \left( \frac{y}{L} \right)^n P_n(\cos \theta)
\end{equation}
Plugging \eqref{one_over_r_expansion} and \eqref{interesting_potential_term} into \eqref{full_potential}, the $n=1$ term cancels, and the constant $n=0$ term can again be neglected. Thus, the effective potential for which the static elastic equation is to be solved becomes
\begin{equation}
\label{effective_potential}
U = - L^2 \left( \sum_{n=2}^\infty \left( \frac{y}{L} \right)^n P_n(\cos \theta) \right)
\end{equation}
Further on, we are going to rename $y$ to $r$ again for simplicity.
%
\subsection{Particular solution}
\label{particular_solution_subsection}
%
We first search a particular solution (i.e. one that doesn't necessarily fit the boundary conditions) for the inhomogeneous elastic equation
\begin{equation}
\label{inhomogeneous_elastic_equation}
u^P_{i,jj} + \frac{\lambda + \mu}{\mu} u^P_{k,ki} = - \frac{1}{\mu} K_i
\end{equation}
If we assume that the force is given by a potential ($K_i = - \partial_i U$), we can find a particular solution by also introducing a potential for the displacement vector: $u^P_i = \frac{1}{\mu} \partial_i \chi$. By inserting this ansatz in \eqref{inhomogeneous_elastic_equation}, we get
\begin{equation}
\frac{\lambda + 2 \mu}{\mu} \partial_i \Delta \chi = \partial_i U
\end{equation}
which can be integrated once to get
\begin{equation}
\label{poisson_equation}
\Delta \chi = \frac{\mu}{\lambda + 2 \mu} U
\end{equation}
In the case the potential $U$ is a harmonic function, there can be found a solution to this equation easily: we can write it as sum of harmonic polynomials:
\begin{equation}
U \left( r, \theta, \phi \right) = \sum_{n=0}^{\infty} U_n
\end{equation}
In the case of rotational symmetry along the $z$-axis, these are proportional to the Legendre polynomials:
\begin{equation}
U_n = E_n r^n P_n \left( \cos \theta \right)
\end{equation}
Since we have
\begin{equation}
\Delta r^2 U_n = U_n \Delta r^2 +r^2 \Delta U_n + 4 x_i \partial_i U_n = 2 \left( 2 n + 3 \right) U_n
\end{equation}
we can write down a solution to equation \eqref{poisson_equation} immediately
\begin{equation}
\chi = \sum_{n=0}^{\infty} \frac{E_n}{2 \left( 2 n + 3 \right)} \frac{\mu}{\lambda + 2 \mu} r^{n+2} P_n \left( \cos \theta \right)
\end{equation}
We thus get
\begin{equation}
\begin{split}
\label{particular_solution}
u^P_r &= \sum_{n=0}^{\infty} \frac{E_n (n + 2)}{2 \left( 2 n + 3 \right)} \frac{1}{\lambda + 2 \mu} r^{n+1} P_n \left( \cos \theta \right) \\
u^P_\theta &= \sum_{n=0}^{\infty} \frac{E_n}{2 \left( 2 n + 3 \right)} \frac{1}{\lambda + 2 \mu} r^{n+1} (- \sin \theta) P'_n \left( \cos \theta \right) \\
u^P_\phi &= 0
\end{split}
\end{equation}
and further (using the abbreviation $k_n = \frac{1}{\lambda + 2 \mu} \frac{E_n}{2 (2 n + 3)} $ and with the Legendre polynomials and their derivatives always understood to be evaluated at $\cos \theta$) for the strain tensor
\begin{equation}
\begin{split}
H^P_{r r} &= \sum_{n=0}^{\infty} k_n (n + 1) (n + 2) r^n P_n \\
H^P_{\theta \theta} &= \sum_{n=0}^{\infty} k_n r^n \left(- \cos \theta P'_n + \sin^2 \theta P''_n + (n + 2) P_n \right) = \\
&= \sum_{n=0}^{\infty} k_n r^n \left(\cos \theta P'_n + (-n^2 + 2) P_n \right) \\
H^P_{\phi \phi} &= \sum_{n=0}^{\infty} k_n r^n \left( - \cos \theta P'_n + (n + 2) P_n \right) \\
H^P_{r \theta} &= \sum_{n=0}^{\infty} k_n r^n (n + 1) (- \sin \theta) P'_n \\
\operatorname{tr} H^P &= \operatorname{div} \mathbf{u}^P = \sum_{n=0}^{\infty} 2 (2 n + 3) k_n r^n P_n
\end{split}
\end{equation}
The other components are zero. For the stress tensor we get
\begin{equation}
\begin{split}
\label{particular_stress_tensor}
\delta \sigma^P_{r r} & = 2 \mu \sum_{n=0}^{\infty} k_n r^n P_n \left( (n + 1) (n + 2) + \frac{\lambda}{\mu} (2 n + 3) \right) \\
\delta \sigma^P_{\theta \theta} & = 2 \mu \sum_{n=0}^{\infty} k_n r^n \left( \cos \theta P'_n + (-n^2 + 2) P_n + \frac{\lambda}{\mu} (2 n + 3) P_n \right) \\
\delta \sigma^P_{\phi \phi} & = 2 \mu \sum_{n=0}^{\infty} k_n r^n \left( - \cos \theta P'_n + (n + 2) P_n + \frac{\lambda}{\mu} (2 n + 3) P_n \right) \\
\delta \sigma^P_{r \theta} & = 2 \mu \sum_{n=0}^{\infty} k_n r^n (n + 1) (- \sin \theta) P'_n
\end{split}
\end{equation}
For the potential \eqref{effective_potential}, we have
\begin{gather}
E_0 = E_1 = 0 \\
E_n = - \rho \omega^2 L^{2-n} \quad (n \ge 2)
\end{gather}
%
%
\subsection{Homogeneous solution}
\label{homogeneous_solution_subsection}
%
The homogeneous elastic equation
\begin{equation}
\mu u^H_{i,jj} + \left( \lambda + \mu \right) u^H_{k,ki} = 0
\end{equation}
is solved by the ansatz from Papkovich and Neuber
\begin{equation}
\label{papkowitsch_neuber_ansatz}
u^H_i = \frac{2 \left(\lambda + 2 \mu \right)}{\lambda + \mu} B_i - \partial_i \left( x_j B_j + B_0 \right)
\end{equation}
where both $B_i$ and $B_0$ are assumed to be harmonic, i.e. they satisfy the Laplace equation. In spherical coordinates $(r, \theta, \phi)$ this becomes
\begin{equation}
\begin{split}
\label{papkowitsch_neuber_ansatz_spherical}
u^H_r &= \frac{2 \left(\lambda + 2 \mu \right)}{\lambda + \mu} B_r - \partial_r \left( r B_r + B_0 \right) \\
u^H_\theta &= \frac{2 \left(\lambda + 2 \mu \right)}{\lambda + \mu} B_\theta - \frac{1}{r} \partial_\theta \left( r B_r + B_0 \right) \\
u^H_\phi &= \frac{2 \left(\lambda + 2 \mu \right)}{\lambda + \mu} B_\phi
\end{split}
\end{equation}
Because of the assumption of rotational symmetry along the $z$-axis, $u^H_i$ and also $B_i$ have to be independant of $\phi$. This is also why $u^H_\phi$ is just proportional to $B_\phi$. In cylindrical coordinates $(\rho, \phi, z)$ the components of $B_i$ are more easy to handle; the connection to the components in spherical coordinates is given by
\begin{gather}
\label{spherical_cylindrical}
B_r = \sin \theta B_\rho + \cos \theta B_z \\
B_\theta = \cos \theta B_\rho - \sin \theta B_z \\
B_\phi = B_\phi
\end{gather}
On the other hand, the connection between the components in cylindrical and cartesian coordinates is given by
\begin{gather}
B_x = \cos \phi B_\rho - \sin \phi B_\phi \\
B_y = \sin \phi B_\rho + \cos \phi B_\phi \\
B_z = B_z
\end{gather}
or, in complex notation
\begin{equation}
B_x + i B_y = \left( B_\rho + i B_\phi \right) e^{i \phi}
\end{equation}
The components of a harmonic vector are harmonic functions only in cartesian coordinates, so $B_z$ is a harmonic function, while $B_\rho$ and $B_\phi$ are not, but $B_\rho e^{i \phi}$ and $B_\phi e^{i \phi}$ are, so they have to be proportional to the spherical harmonics with $m=1$, and so $B_\rho$ and $B_\phi$ are proportional to the associated Legendre polynomials with $m=1$. Thus, they can be expressed by
\begin{gather}
\label{harmonical}
B_\rho = \sum_{n=0}^{\infty} A_n r^n (- \sin \theta) P'_n \left( \cos \theta \right) \\
B_\phi = \sum_{n=0}^{\infty} C^*_n r^n (- \sin \theta) P'_n \left( \cos \theta \right) \\
B_z = \sum_{n=0}^{\infty} A^*_n r^n P_n \left( \cos \theta \right)
\end{gather}
inserting \eqref{harmonical} into \eqref{spherical_cylindrical} yields
\begin{gather}
B_r = \sum_{n=0}^{\infty} - A_n r^n P'_n (\cos \theta) \sin^2 \theta + A^*_n r^n P_n(\cos \theta) \cos \theta \\
B_\theta = \sum_{n=0}^{\infty} - r^n A_n P'_n(\cos \theta) \sin \theta \cos \theta - A^*_n r^n P_n(\cos \theta) \sin \theta
\end{gather}
using the relations (see e.g. \cite{whittaker-watson})
\begin{gather}
\left( x^2 - 1 \right) P'_n(x) = n x P_n(x) - n P_{n-1}(x) \\
x P'_n(x) = n P_n(x) + P'_{n-1}(x)
\end{gather}
this becomes
\begin{gather}
B_r = A_0^* \cos \theta + \sum_{n=1}^{\infty} r^n \left( - A_n n P_{n-1} + \cos \theta P_n \left( A_n n + A^*_n \right) \right) \\
B_\theta = -A_0^* \sin \theta + \sum_{n=1}^{\infty} - r^n \sin \theta \left( A_n P'_{n-1} + P_n \left( A_n n + A^*_n \right) \right)
\end{gather}
One of the four harmonic functions in this ansatz can be chosen arbitrarily, so we can significantly simplify everything by setting
\begin{equation}
A_n n + A^*_n = 0
\end{equation}
using this, performing an index-shift and renaming the $A_{n+1}$ back to $A_n$, we get
\begin{gather}
B_r = \sum_{n=0}^{\infty} - A_n (n+1) r^{n+1} P_n (\cos \theta) \\
B_\theta = \sum_{n=0}^{\infty} - A_n r^{n+1} \sin \theta P_n'(\cos \theta)
\end{gather}
for the fourth harmonic function $B_0$ we use
\begin{equation}
B_0 = \sum_{n=0}^{\infty} - B_n r^n P_n(\cos \theta)
\end{equation}
plugging all this into the ansatz \eqref{papkowitsch_neuber_ansatz_spherical}, we get
\begin{equation}
\begin{split}
\label{homogeneous_solution}
u^H_r &= \sum_{n=0}^{\infty} \left( A_n (n + 1) \left( n - \frac{2 \mu}{\lambda + \mu} \right) r^{n+1} + B_n n r^{n-1} \right) P_n(\cos \theta) \\
u^H_\theta &= \sum_{n=0}^{\infty} \left( A_n \left( n + \frac{3 \lambda + 5 \mu}{\lambda + \mu} \right) r^{n+1} + B_n r^{n-1} \right) (-\sin \theta) P'_n(\cos \theta) \\
u^H_\phi &= \sum_{n=0}^{\infty} C_n r^n (- \sin \theta) P'_n \left( \cos \theta \right)
\end{split}
\end{equation}
this gives the strain tensor
\begin{equation}
\begin{split}
H^H_{r r} &= \sum_{n=0}^{\infty} \left( A_n (n + 1)^2 \left( n - \frac{2 \mu}{\lambda + \mu} \right) r^n + B_n n (n - 1) r^{n-2} \right) P_n \\
H^H_{\theta \theta} &= \sum_{n=0}^{\infty} \left( - \left( A_n \left( n^2 + n \frac{2 \lambda + 4 \mu}{\lambda + \mu} + \frac{2 \mu}{\lambda + \mu} \right) \left( n + 1 \right) r^n + B_n n^2 r^{n-2} \right) P_n \right) + \\
& \hphantom{= \sum_{n=0}^{\infty}} \quad + \left( A_n \left( n + \frac{3 \lambda + 5 \mu}{\lambda + \mu} \right) r^n + B_n r^{n-2} \right)  \cos \theta P_n' \\
H^H_{\phi \phi} &= \sum_{n=0}^{\infty} \left( A_n (n + 1) \left( n - \frac{2 \mu}{\lambda + \mu} \right) r^n + B_n n r^{n-2} \right) P_n + \\
& \hphantom{= \sum_{n=0}^{\infty}} \quad + \left( A_n \left( n + \frac{3 \lambda + 5 \mu}{\lambda + \mu} \right) r^n + B_n r^{n-2} \right) (- \cos \theta) P'_n \\
H^H_{r \theta} &= \sum_{n=0}^{\infty} \left( A_n \left( n^2 + 2 n - \frac{\mu}{\lambda + \mu} \right) r^n + B_n (n - 1) r^{n-2} \right) (- \sin \theta) P'_n \\
H^H_{r \phi} &= \frac{1}{2} \sum_{n=0}^{\infty} \left( (n-1) C_n r^{n-1} \right) (- \sin \theta) P'_n \\
H^H_{\theta \phi} &= \sum_{n=0}^{\infty} C_n r^{n-1} \left( \cos \theta P'_n - \frac{1}{2} n (n+1) P_n \right) \\
\operatorname{tr} H^H &= \operatorname{div} \mathbf{u}^H = \sum_{n=0}^{\infty} - \frac{2 \mu}{\lambda + \mu} A_n (2 n + 3) (n + 1) r^n P_n
\end{split}
\end{equation}
which in turn gives the stress tensor
\begin{equation}
\begin{split}
\delta \sigma^H_{r r} & = 2 \mu \sum_{n=0}^{\infty} \left( A_n (n + 1) \left( n^2 - n - \frac{3 \lambda + 2 \mu}{\lambda + \mu} \right) r^n + B_n n (n - 1) r^{n-2} \right) P_n \\
\delta \sigma^H_{\theta \theta} & = 2 \mu \sum_{n=0}^{\infty} \left( - \left( A_n \left( n^2 + 4 n + \frac{3 \lambda + 2 \mu}{\lambda + \mu} \right) \left( n + 1 \right) r^n + B_n n^2 r^{n-2} \right) P_n \right) + \\
& \hphantom{= 2 \mu \sum_{n=0}^{\infty}} \quad + \left( A_n \left( n + \frac{3 \lambda + 5 \mu}{\lambda + \mu} \right) r^n + B_n r^{n-2} \right)  \cos \theta P_n' \\
\delta \sigma^H_{\phi \phi} &= 2 \mu \sum_{n=0}^{\infty} \left( A_n (n + 1) \left( n \frac{\mu - \lambda}{\lambda + \mu} - \frac{3 \lambda + 2 \mu}{\lambda + \mu} \right) r^n + B_n n r^{n-2} \right) P_n + \\
& \hphantom{= 2 \mu \sum_{n=0}^{\infty}} \quad + \left( A_n \left( n + \frac{3 \lambda + 5 \mu}{\lambda + \mu} \right) r^n + B_n r^{n-2} \right) (- \cos \theta) P'_n \\
\delta \sigma^H_{r \theta} & = 2 \mu \sum_{n=0}^{\infty} \left( A_n \left( n^2 + 2 n - \frac{\mu}{\lambda + \mu} \right) r^n + B_n (n - 1) r^{n-2} \right) (- \sin \theta) P'_n \\
\delta \sigma^H_{r \phi} &= \mu \sum_{n=0}^{\infty} \left( (n-1) C_n r^{n-1} \right) (- \sin \theta) P'_n \\
\delta \sigma^H_{\theta \phi} &= 2 \mu \sum_{n=0}^{\infty} C_n r^{n-1} \left( \cos \theta P'_n - \frac{1}{2} n (n+1) P_n \right) \\
\end{split}
\end{equation}
To get the overall solution, one has to add the particular solution \eqref{particular_solution} and the homogeneous solution \eqref{homogeneous_solution} and submit it to the boundary condition to determine the constants $A_n$, $B_n$ and $C_n$:
\begin{equation}
\begin{split}
\label{boundary_conditions}
0 &= \delta \sigma^H_{r r}|_{r=a} + \delta \sigma^P_{r r}|_{r=a} \\
0 &= \delta \sigma^H_{r \theta}|_{r=a} + \delta \sigma^P_{r \theta}|_{r=a} \\
0 &= \delta \sigma^H_{r \phi}|_{r=a}
\end{split}
\end{equation}
The equation for $\delta \sigma_{r \phi}$ becomes
\begin{equation}
(n-1) C_n = 0
\end{equation}
so the $C_n$ for $n \ne 0$ vanish, while $C_1$ is arbitrary. The corresponding displacement
\begin{equation}
u_\phi = - C_1 r \sin \theta
\end{equation}
corresponds to rigid rotations along the $z$-axis and can therefore safely be neglected.

The boundary conditions for $\delta \sigma_{r r}$ and $\delta \sigma_{r \theta}$ become a linear $2 \times 2$ system for the variables $A_n$ and $B_n$
\begin{gather}
\label{linear_system_for_A_B}
\left( n^2 + 2 n - \frac{\mu}{\lambda + \mu} \right) a^n A_n + (n - 1) a^{n-2} B_n = - k_n a^n (n + 1) \\
(n + 1) \left( n^2 - n - \frac{3 \lambda + 2 \mu}{\lambda + \mu} \right) a^n A_n + n (n - 1) a^{n-2} B_n = \\
= - k_n a^n \left( (n+1)(n+2) + \frac{\lambda}{\mu} (2 n + 3) \right)
\end{gather}
with the determinant
\begin{equation}
D_n = (n - 1) \left( 2 n (n - 1) + \frac{3 \lambda + 2 \mu}{\lambda + \mu} (2 n + 1) \right) a^{2n-2}
\end{equation}
Because of $\frac{3 \lambda + 2 \mu}{\lambda + \mu} > 0$, we have $D_0 < 0$, $D_1 = 0$ and $D_n > 0$ for $n \ge 2$. Because $k_0 = k_1 = 0$, this gives $A_0 = B_0 = A_1 = 0$, while $B_1$ is arbitrary. The corresponding displacements
\begin{equation}
\begin{split}
u_r &= B_1 \cos \theta \\
u_\theta &= - B_1 \sin \theta
\end{split}
\end{equation}
are rigid translations along the $z$-axis, lying in the kernel of the linearized elasticity operator. Thus, the value of $B_1$ has to be calculated from the equilibrium conditions after the rest of the solution has been determined.

For $n \ge 2$, we get the following unique solution for the system \eqref{linear_system_for_A_B}:
\begin{equation}
\begin{split}
A_n &= \frac{1}{D_n} k_n a^{2n-2} (n - 1) \left( 2 (n + 1) + \frac{\lambda}{\mu} (2 n + 3) \right) \\
B_n &= - \frac{1}{D_n} k_n a^{2n} \frac{\lambda + 2 \mu}{\lambda + \mu} n (2 n + 3) \left( (n+1) + \frac{\lambda}{\mu} (n + 2) \right)
\end{split}
\end{equation}
combining everything gives the solution of the elastic equations
\begin{equation}
\begin{split}
\label{whole_solution}
u_r &= s \sum_{n=2}^\infty \left( F_n \left( \frac{r}{L} \right)^{n+1} + G_n \frac{a^2}{L^2} \left( \frac{r}{L} \right)^{n-1} \right) P_n(\cos \theta) \\
u_\theta &= s \sum_{n=2}^\infty \left( H_n \left( \frac{r}{L} \right)^{n+1} + I_n \frac{a^2}{L^2} \left( \frac{r}{L} \right)^{n-1} \right) \frac {d P_n}{d \theta}(\cos \theta) \\
u_\phi &= 0
\end{split}
\end{equation}
with
\begin{equation}
\begin{split}
s   &= \frac{\rho \omega^2 L^3}{2 \left( \lambda + \mu \right)} = \frac{\rho G M}{2 \left( \lambda + \mu \right)} \\
F_n &= - \frac{n \left( n + \frac{\lambda}{\mu} (n + 1) \right)}{2 n (n - 1) + \frac{3 \lambda + 2 \mu}{\lambda + \mu} (2n + 1)} \\
G_n &= n I_n \\
H_n &= - \frac{n + 2 + \frac{\lambda}{\mu} (n + 3)}{2 n (n - 1) + \frac{3 \lambda + 2 \mu}{\lambda + \mu} (2n + 1)} \\
I_n &= \frac{n \left( n + 1 + \frac{\lambda}{\mu} (n + 2) \right)}{(n - 1) \left( 2 n (n - 1) + \frac{3 \lambda + 2 \mu}{\lambda + \mu} (2n + 1) \right)}
\end{split}
\end{equation}
%
%
%
%
%
It is noteworthy that the Lam\'e constants only occur in ratios except in the scale factor $s$.

The overall solution \eqref{whole_solution} has the strain tensor
\begin{equation}
\begin{split}
H_{r r} &= \frac{s}{L} \sum_{n=2}^{\infty} \left( (n + 1) F_n \left( \frac{r}{L} \right)^n + (n - 1) G_n \frac{a^2}{L^2} \left( \frac{r}{L} \right)^{n-2} \right) P_n(\cos \theta) \\
H_{\theta \theta} &= \frac{s}{L} \sum_{n=2}^{\infty} \left( (F_n - n (n+1) H_n) \left( \frac{r}{L} \right)^n - n^2 I_n \frac{a^2}{L^2} \left( \frac{r}{L} \right)^{n-2} \right) P_n(\cos \theta) + \\
& \qquad + \left( H_n \left( \frac{r}{L} \right)^n + I_n \frac{a^2}{L^2} \left( \frac{r}{L} \right)^{n-2} \right) \cos \theta P_n'(\cos \theta) \\
H_{\phi \phi} &= \frac{s}{L} \sum_{n=2}^{\infty} \left( F_n \left( \frac{r}{L} \right)^n + G_n \frac{a^2}{L^2} \left( \frac{r}{L} \right)^{n-2} \right) P_n(\cos \theta) + \\
& \qquad + \left( H_n \left( \frac{r}{L} \right)^n + I_n \frac{a^2}{L^2} \left( \frac{r}{L} \right)^{n-2} \right) (- \cos \theta) P_n'(\cos \theta) \\
H_{r \theta} &= \frac{s}{L} \sum_{n=2}^{\infty} \left( \frac{1}{2} \left( F_n + n H_n \right) \left( \frac{r}{L} \right)^n + (n - 1) I_n \frac{a^2}{L^2} \left( \frac{r}{L} \right)^{n-2} \right) (-\sin \theta) P'_n \\
\operatorname{tr} H &= \frac{s}{L} \sum_{n=2}^{\infty} \left( (n + 3) F_n - n (n + 1) H_n \right) \left( \frac{r}{L} \right)^n P_n(\cos \theta)
\end{split}
\end{equation}
From that it is easy to check that the boundary conditions \eqref{boundary_conditions} are indeed satisfied using the algebraic identities
\begin{gather}
F_n + n H_n + 2 (n-1) I_n = 0 \\
2 \mu \left( (n+1) F_n + (n-1) G_n \right) + \lambda \left( (n+3) F_n - 3 n (n+1) H_n \right) = 0
\end{gather}
Also the elastic equations can be checked:
\begin{equation}
\begin{split}
\left( \Delta \vec u \right)_r &= \frac{1}{r^2} \frac{\partial}{\partial r} \left( r^2 \frac{\partial u_r}{\partial r} \right) + \frac{1}{r^2} \frac{\partial^2 u_r}{\partial \theta^2} + \frac{1}{r^2} \frac{\cos \theta}{\sin \theta} \frac{\partial u_r}{\partial \theta} - \frac{2 u_r}{r^2} - \frac{2}{r^2 \sin \theta} \frac{\partial (u_\theta \sin \theta)}{\partial \theta} = \\
&= \frac{s}{L^2} \sum_{n=2}^\infty \left( 2 n F_n + 2 n (n + 1) H_n \right) \left( \frac{r}{L} \right)^{n-1} P_n(\cos \theta)  \\
\left( \Delta \vec u \right)_\theta &= \frac{1}{r^2} \frac{\partial}{\partial r} \left( r^2 \frac{\partial u_\theta}{\partial r} \right) + \frac{1}{r^2} \frac{\partial^2 u_\theta}{\partial \theta^2} + \frac{1}{r^2} \frac{\cos \theta}{\sin \theta} \frac{\partial u_\theta}{\partial \theta} - \frac{u_\theta}{r^2 \sin^2 \theta} + \frac{2}{r^2} \frac{\partial u_r}{\partial \theta} = \\
&= \frac{s}{L^2} \sum_{n=2}^\infty \left( 2 F_n + 2 (n + 1) H_n \right) \left( \frac{r}{L} \right)^{n-1} \frac{d P_n}{d \theta}(\cos \theta)  \\
\left( \Delta \vec u \right)_\phi &= 0
\end{split}
\end{equation}
\begin{equation}
\begin{split}
\left( \operatorname{grad} \operatorname{div} \vec u \right)_r &= \frac{s}{L^2} \sum_{n=2}^\infty \left( n (n + 3) F_n - n^2 (n+1) H_n \right) \left( \frac{r}{L} \right)^{n-1} P_n(\cos \theta)  \\
\left( \operatorname{grad} \operatorname{div} \vec u \right)_\theta &= \frac{s}{L^2} \sum_{n=2}^\infty \left( (n + 3) F_n - n (n+1) H_n \right) \left( \frac{r}{L} \right)^{n-1} \frac{d P_n}{d \theta}(\cos \theta) \\
\left( \operatorname{grad} \operatorname{div} \vec u \right)_\phi &= 0 \\
\end{split}
\end{equation}
From the algebraic identity
\begin{equation}
\mu \left( 2 F_n + 2 (n+1) H_n \right) + (\lambda + \mu) \left( (n+3) F_n - n (n+1) H_n \right) = - 2 (\lambda + \mu)
\end{equation}
it then follows that the solution \eqref{whole_solution} indeed satisfies the elastic equations with the potential \eqref{effective_potential}.

%
\subsection{Displacement}
%
To complete our solution, we still have to calculate the displacement factor $C(t)$ of the Killing part of the solution (compare equation \eqref{phi_kerne_decomposition_special_case}). We do this in first order in the linearization parameter $t=\omega^2$ as well, using the explicit formula the implicit function theorem provides us with on the equilibrium condition \eqref{equilibrium_condition_z}:
\begin{equation}
C \cong \left. \frac{d C}{d t}\right|_{t=0} t = - \frac{\frac{\partial N}{\partial t}(0, 0) \omega^2}{\frac{\partial N}{\partial C}(0, 0)}
\end{equation}
The denominator has already been calculated in equation \eqref{N_by_C}. It remains to compute
\begin{equation}
\label{N_partial_lambda}
\frac{\partial N}{\partial t} (0, 0) t = \int_\mathcal{B} \left( \left( K_{i} n^i \right)_{,j} u^j \right) d^3 X
\end{equation}
with $K_i$ being the force field, divided by the linearization constant:
\begin{equation}
K_i = \partial_i \left( \frac{1}{2} \left( \mathbf{x}^2 - \left( \mathbf{x}, \mathbf{m} \right)^2 \right) + \frac{L^3}{r} \right)
\end{equation}
The centrifugal term in $\left( K_{i} n^i \right)_{,j}$ is just proportional to $n_j = P_1 (\cos \theta) e^r_j + P_1^{(1)} (\cos \theta) e^\theta_j$, so it doesn't contribute to the integral because of the orthogonality conditions for the (associated) Legendre polynomials (see equation \eqref{legendre_orthogonality} below), since the sums for $u^j$ only start with $l=2$. So it remains to calculate the contribution from
\begin{equation}
\frac{L^3}{\sqrt{|\mathbf{y}|^2 - 2 L |\mathbf{y}| \cos \theta + L^2}} = L^2 \sum_{l=0}^\infty \left( \frac{|\mathbf{y}|}{L} \right)^l P_l(\cos \theta)
\end{equation}
From differentiating with resprect to $y^i$, we get
\begin{equation}
\label{one_over_r_term_differentiated}
\partial_i \frac{L^3}{\sqrt{|\mathbf{y}|^2 - 2 L |\mathbf{y}| \cos \theta + L^2}} = L \sum_{l=1}^\infty \left( \frac{|\mathbf{y}|}{L} \right)^{l-1} \left( l P_l(\cos \theta) e^r_i + (- \sin \theta) P_l'(\cos \theta) e^\theta_i \right)
\end{equation}
The translational Killing vector in direction of the gravitational center is
\begin{equation}
\xi^i = e_z^i = \cos \theta e_r^i - \sin \theta e_\theta^i
\end{equation}
Inserting this in \eqref{one_over_r_term_differentiated} yields
\begin{equation}
\partial_i \frac{L^3}{\sqrt{|\mathbf{y}|^2 - 2 L |\mathbf{y}| \cos \theta + L^2}} \xi^i = L \sum_{l=1}^\infty \left( \frac{r}{L} \right)^{l-1} \left( l \cos \theta P_l(\cos \theta) + \sin^2 \theta P_l'(\cos \theta) \right)
\end{equation}
Now using the Legendre polynomial relation
\begin{equation}
(x^2 - 1) P_l'(x) = l x P_l(x) - l P_{l-1}
\end{equation}
this becomes (after an index-shift)
\begin{equation}
\partial_i \frac{L^3}{\sqrt{|\mathbf{y}|^2 - 2 L |\mathbf{y}| \cos \theta + L^2}} \xi^i = L \sum_{l=0}^\infty (l+1) \left( \frac{r}{L} \right)^l P_l(\cos \theta)
\end{equation}
which becomes, after a further derivative
\begin{equation}
\begin{split}
\partial_j \left( \partial_i \frac{L^3}{\sqrt{|\mathbf{y}|^2 - 2 L |\mathbf{y}| \cos \theta + L^2}} \xi^i \right) u^j &= \sum_{l=0}^\infty \left( l (l+1) \left( \frac{r}{L} \right)^{l-1} P_l(\cos \theta) u_r + \right. \\
& \qquad + \left. (l+1) \left( \frac{r}{L} \right)^{l-1} \frac{d P_l}{d \theta}(\cos \theta) u_\theta \right)
\end{split}
\end{equation}
This can be easily integrated over $\mathcal{B}$, using the following orthogonality conditions for the (associated) Legendre polynomials (see \cite{whittaker-watson})
\begin{equation}
\label{legendre_orthogonality}
\begin{split}
\int_{-1}^1 P_l(x) P_n(x) dx = \frac{2}{2 n + 1} \delta_{l n} \\
\int_{-1}^1 P^{(1)}_l(x) P^{(1)}_n(x) dx = \frac{2 n (n+1)}{2 n + 1} \delta_{l n}
\end{split}
\end{equation}
the integral \eqref{N_partial_lambda} then becomes
\begin{equation}
\frac{\partial N}{\partial t} (0, 0) t = 4 \pi a^3 s \sum_{n=2}^\infty \left( \frac{a}{L} \right)^{2 n} \frac{n (n+1)}{2n+1} \left( \frac{F_n + (n+1) H_n}{2 n + 3} + I_n \right)
\end{equation}

Thus, for small values of the linearization parameter $t = \omega^2$, we get the linear approximation
\begin{equation}
\label{displacement_value}
\begin{split}
C & \cong \left. \frac{d C}{d t}\right|_{t=0} t = - \frac{\frac{\partial N}{\partial t}(0, 0) \omega^2}{\frac{\partial N}{\partial C}(0, 0)} = \\
&= - s \sum_{n=2}^\infty \left( \frac{a}{L} \right)^{2 n} \frac{n (n+1)}{2n+1} \left( \frac{F_n + (n+1) H_n}{2 n + 3} + I_n \right)
\end{split}
\end{equation}
Using the formulas for $F_n$, $H_n$ and $I_n$, it is easy to see that each summand in \eqref{displacement_value} is positive, i.e. the overall minus sign makes $C$ negative.

\begin{figure}[hp]
\includegraphics{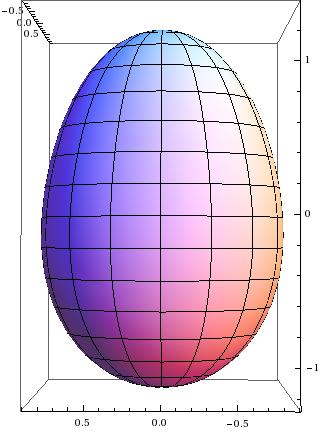}
\caption{A graphic display of the solution \eqref{whole_solution}. The material used is steel ($\lambda$ = 108 GPa, $\mu$ = 78 GPa, $\rho_0$ = 7860 kg/$m^3$). For the orbit velocity $\omega L$ = 9.3 km/s was used, and $L=3a$. The gravitational centre is in the direction of the top side of the picture.}
\label{nichtrelativistische_graphik}
\end{figure}

\clearpage

%
\section{Relativistic case}
%
We now want to consider an elastic sphere on a circular orbit in the Schwarzschild metric
\begin{equation}
\label{schwarzschild}
g_{\mu \nu} dx^\mu dx^\nu = - \left( 1 - \frac{2 G M}{c^2 r} \right) c^2 dt^2 + \left( 1 - \frac{2 G M}{c^2 r} \right)^{-1} dr^2 + r^2 \left( d \theta^2 + \sin^2 \theta d \varphi^2 \right)
\end{equation}
We first recapitulate that these orbits (for point particles) can be found using the following neat property: we assume $\xi^\mu$ to be a timelike Killing vector field, i.e. it satisfies the Killing equation
\begin{equation}
\label{killing_equation}
\nabla_\mu \xi_\nu + \nabla_\nu \xi_\mu = 0
\end{equation}
Then a orbit $\gamma$ of the flow of $\xi^\mu$ with the normalized tangent vector (4-velocity)
\begin{equation}
\label{xi_normalization}
u_\mu = \frac{\xi_\mu}{\sqrt{- \xi_\lambda \xi_\rho g^{\lambda \rho}}}
\end{equation}
is geodesic, i.e. satisfies
\begin{equation}
\label{geodesic_equation}
u^\mu \nabla_\mu u_\nu = 0
\end{equation}
if and only if the gradient of $\xi^2$ vanishes everywhere on $\gamma$:
\begin{equation}
\nabla_\mu (\xi_\lambda \xi_\rho g^{\lambda \rho}) = 2 \xi^\lambda \nabla_\mu \xi_\lambda = 0
\end{equation}
This can be easily seen by inserting \eqref{xi_normalization} into \eqref{geodesic_equation} to get
\begin{equation}
\frac{\xi^\mu}{\sqrt{- \xi^2}} \nabla_\mu \frac{\xi_\nu}{\sqrt{- \xi^2}} = - \frac{1}{2} \frac{1}{(-\xi^2)} \nabla_\nu \xi^2 + \frac{1}{(-\xi^2)^2} \xi^\mu \xi^\lambda \nabla_\mu \xi_\lambda
\end{equation}
and noting that the second term vanishes identically because of the Killing equation \eqref{killing_equation}.

We now apply this to the helical Killing vector (remember that constant linear combinations of Killing vectors are Killing vectors themselves)
\begin{equation}
\label{helical_killing_vector}
\xi^\mu \partial_\mu = \partial_t + \omega \partial_\varphi
\end{equation}
of the Schwarzschild metric \eqref{schwarzschild}. For its normal value, we get
\begin{equation}
\label{xi_squared}
\xi^\lambda \xi^\rho g_{\lambda \rho} = - c^2 \left( 1 - \frac{2 G M}{c^2 r} \right) + \omega^2 r^2 \sin^2 \theta
\end{equation}
From forming the gradient of \eqref{xi_squared}, we get the equations
\begin{gather}
\label{geodesic_condition_1}
- \frac{2 G M}{r^2} + 2 \omega^2 r \sin^2 \theta = 0 \\
\label{geodesic_condition_2}
2 \omega^2 r^2 \sin \theta \cos \theta = \omega^2 r^2 \sin 2 \theta = 0
\end{gather}
Equation \eqref{geodesic_condition_2} has the solutions $\theta = 0$, $\theta = \pi / 2$ and $\theta=\pi$, but equation \eqref{geodesic_condition_1} can only be solved for $\theta = \pi / 2$. We thus got the circular geodesics in the equatorial plane with $r = L = const$, where $\omega$, $M$ and $L$ have to satisfy the same relation as in the Newtonian case:
\begin{equation}
\omega^2 L = \frac{G M}{L^2}
\end{equation}
In order for $\xi_\mu$ to be timelike, i.e. equation \eqref{xi_squared} to be negative, it is necessary that $L > 3 G M / c^2$.

We now consider an elastic sphere moving on such an orbit. We do so by changin to the co-rotating coordinate system. We do this by introducing co-rotating coordinates:
\begin{equation}
\varphi' = \varphi + \omega t
\end{equation}
Using this coordinate transformation (and renaming $\varphi'$ back to $\varphi$ afterwards), \eqref{schwarzschild} becomes
\begin{equation}
\begin{split}
\label{corotating_schwarzschild}
g_{\mu \nu} dx^\mu dx^\nu = & - \left( 1 - \frac{2 G M}{c^2 r} - \frac{\omega^2 r^2 \sin^2 \theta}{c^2} \right) c^2 dt^2 + \left( 1 - \frac{2 G M}{c^2 r} \right)^{-1} dr^2 + \\
& + 2 \omega r^2 \sin^2 \theta d \varphi dt + r^2 \left( d \theta^2 + \sin^2 \theta d \varphi^2 \right)
\end{split}
\end{equation}
We introduce the potential term
\begin{equation}
e^\frac{2 U}{c^2} = 1 - \frac{2 G M}{c^2 r} - \frac{\omega^2 r^2 \sin^2 \theta}{c^2}
\end{equation}
The metric $h_{i j}$ on $N$, the quotient of the Schwarzschild spacetime $M$ along the helical Killing vector $\xi^\mu$ given by \eqref{helical_killing_vector}, is then given by (see \cite{beig-schmidt-1})
\begin{equation}
h_{i j} = g_{i j} - g_{0 i} g_{0 j} / g_{0 0}
\end{equation}
With \eqref{corotating_schwarzschild}, we get
\begin{equation}
\label{three-metric}
h_{i j} dx^i dx^j = \left( 1 - \frac{2 G M}{c^2 r} \right)^{-1} dr^2 + r^2 d \theta^2 + \left( r^2 \sin^2 \theta + \frac{\omega^2}{c^2} r^4 \sin^4 \theta e^{-\frac{2 U}{c^2}} \right) d \varphi^2
\end{equation}
The elastostatic equations are
\begin{equation}
\label{nonlinear_elastostatic_equation}
e^{-\frac{U}{c^2}} \nabla_A \left( e^\frac{U}{c^2} \sigma_i{}^A \right) - \left( 1 + \frac{w}{c^2} \right) \partial_ i U = 0
\end{equation}
with
\begin{equation}
\label{definition_nabla}
\nabla_A \sigma_i{}^A = V^{-1} \partial_A \left( V \sigma_i{}^A \right) - \Gamma^k_{i j} \sigma_k{}^A \Phi^j_A
\end{equation}
The Christoffel symbols of the metric \eqref{three-metric} become
\begin{equation}
\begin{split}
&\Gamma^r{}_{r r} = \frac{1}{2} h_{rr,r}/h_{rr} = - \frac{\omega^2 L^3}{c^2 r^2} \left( 1 - \frac{2 \omega^2 L^3}{c^2 r} \right)^{-1} \\
&\Gamma^r{}_{\theta \theta} = - \frac{1}{2} h_{\theta \theta,r} / h_{rr} = -r + \frac{2 \omega^2 L^3}{c^2} \\
&\Gamma^r{}_{\varphi \varphi} = - \frac{1}{2} h_{\varphi \varphi, r} / h_{r r} = -r \sin^2 \theta + \mathcal{O} \left(\frac{\omega^2}{c^2} \right) \\
&\Gamma^\theta{}_{r \theta} = \frac{1}{2} h_{\theta \theta,r} / h_{\theta \theta} =  \frac{1}{r} \\
&\Gamma^\theta{}_{\varphi \varphi} = - \frac{1}{2} h_{\varphi \varphi, \theta} / h_{\theta \theta} = - \sin \theta \cos \theta + \mathcal{O} \left( \frac{\omega^2}{c^2} \right) \\
&\Gamma^\varphi{}_{r \varphi} = \frac{1}{2} h_{\varphi \varphi, r} / h_{\varphi \varphi} = \frac{1}{r} + \mathcal{O} \left( \frac{\omega^2}{c^2} \right) \\
&\Gamma^\varphi{}_{\theta \varphi} = \frac{1}{2} h_{\varphi \varphi, \theta} / h_{\varphi \varphi} = \cot \theta + \mathcal{O} \left( \frac{\omega^2}{c^2} \right)
\end{split}
\end{equation}
The not mentioned ones are zero. For the last two Christoffel symbols we have used the Taylor expansion
\begin{equation}
h^{\varphi \varphi} = \frac{1}{h_{\varphi \varphi}} = \frac{1}{r^2 \sin^2 \theta} \left( 1 - \frac{\omega^2}{c^2} r^2 \sin^2 \theta e^{-2 U / c^2} + \mathcal{O} \left( \frac{\omega^4}{c^4} \right) \right)
\end{equation}
We notice that the Christoffel symbols consist of two parts: the regular Christoffel symbols of flat space in spherical coordinates (which we will denote $\mathring \Gamma^i{}_{j k}$ from now), plus some correction terms proportional to $\omega^2/c^2$, which we will denote $\widetilde \Gamma^i{}_{j k}$.

Like in \cite{beig-schmidt-1}, we treat the relativistic equation by splitting it in the nonrelativistic problem plus some correction terms. Thus we write
\begin{equation}
e^{U/c^2} \sigma_i{}^A = \mathring \sigma_i{}^A + \omega^2 \widetilde \sigma_i{}^A
\end{equation}
Inserting this decomposition and \eqref{definition_nabla} in \eqref{nonlinear_elastostatic_equation} and multiplying with $e^{U/c^2}$ yields
\begin{equation}
\label{separated_elastostatic_equation}
\frac{1}{V} \partial_A \left( V \mathring \sigma_i{}^A \right) + \omega^2 \frac{1}{V} \partial_A \left( V \widetilde \sigma_i{}^A \right) - e^{U/c^2} \Gamma^k{}_{i j} \phi^j{}_A \sigma_k{}^A - e^{U/c^2} \left( 1 + \frac{w}{c^2} \right) \partial_ i U = 0
\end{equation}
The linearized equation can then be obtained by differentiating with respect to $\omega^2$ and setting it zero afterwards. For simplicity, we perturb around an relaxed configuration, i.e. we assume
\begin{equation}
\label{relaxed_background}
\left. \mathring \sigma_i{}^A \right|_{\left( \omega^2=0, \phi = \bar \phi \right)} = 0
\end{equation}
As usual, we will identify $\mathcal{B}$ with the part of physical space N occupied by the body in the reference configuration, i.e.
\begin{equation}
\label{body_space_identification}
\mathring f^A(x^i) = \delta^A_i x^i
\end{equation}
The superscript ?? will be used from now on to denote quantities referring to the reference configuration only. We also assume the stored energy function $w$ to be zero in the reference configuration
\begin{equation}
\label{w_zero_in_reference_configuration}
\left. w(H^{AB}, X) \right|_{(\omega^2 = 0, \phi = \bar \phi)} = 0
\end{equation}
this is necessary, because $w$ explicitely occurs in the relativistic elastostatic equations (as opposed to the nonrelativistic ones), and thus they are not invariant under adding a constant to $w$.

The perturbation of $\mathring \sigma_i{}^A$ is then just equal to the standard expression well-known from nonrelativistic elasticity theory:
\begin{equation}
\label{nonrelativistic_linear_stress_tensor}
\delta \mathring \sigma_i{}^A = -2 \mathring H^{A B} \mathring f^C{}_i \mathring L_{BCDE} \delta H^{DE}
\end{equation}
where the $L_{BCDE}$ are given by the Lame constants:
\begin{equation}
\label{definition_lame}
\mathring L_{BCDE} = \frac{1}{4} \left( \lambda \delta_{BC} \delta_{DE} + \mu \left( \delta_{BD} \delta_{CE} + \delta_{BE} \delta_{CD} \right) \right)
\end{equation}
The force term just becomes the nonrelativistic one in the perturbation process (remember that $w=0$ and $e^{U/c^2}=1$ in the reference configuration):
\begin{equation}
\label{force_density}
\left. \frac{d}{d \left( \omega^2 \right)} \left( - e^{U/c^2} \left( 1 + \frac{w}{c^2} \right) \partial_ i U \right) \right|_{\omega^2 = 0} = - \frac{L^3}{r^2} \frac{x_i}{r} + \left( \begin{array}{c} x \\ y \\ 0 \end{array} \right)_i =: f_i
\end{equation}
For the Christoffel symbol term in \eqref{separated_elastostatic_equation}, we get
\begin{equation}
\left. \frac{d}{d \left( \omega^2 \right)} \left( - e^{U/c^2} \Gamma^k{}_{i j} \phi^j{}_A \sigma_k{}^A \right) \right|_{\omega^2 = 0} = - \mathring \Gamma^k{}_{i j} \delta^j_A \left( \delta \mathring \sigma_i{}^A + \left. \widetilde \sigma_i{}^A \right|_{\omega^2=0} \right)
\end{equation}
The Christoffel symbols $\mathring \Gamma^k{}_{i j}$ together with the partial derivatives form the regular affine connection on flat space. When we denote this $\mathring \nabla_A$, we get for the perturbed elastostatic equation
\begin{equation}
\label{linearized_elastic_equation}
\mathring \nabla_A \delta \mathring \sigma_i{}^A + \mathring \nabla_A \left. \widetilde \sigma_i{}^A \right|_{\omega^2=0} + f_i = 0
\end{equation}
The first term is the nonrelativistic elasticity operator, which becomes with \eqref{nonrelativistic_linear_stress_tensor} and \eqref{definition_lame}
\begin{equation}
\mathring \nabla_A \delta \mathring \sigma_i{}^A = \mu \Delta \delta \phi^i + \left( \lambda + \mu \right) \operatorname{grad} \operatorname{div} \delta \phi^i
\end{equation}
The perturbations $\delta \phi^i$ are proportional to the displacement vector $u^i$:
\begin{equation}
u^i(x^j) = \omega^2 \delta \phi^i( \mathring f^A(x^i))
\end{equation}
The perturbations of the correction term of the stress tensor is given by
\begin{equation}
\label{correction_definition}
\left. \widetilde \sigma_i{}^A \right|_{\omega^2=0} = \lim_{\omega^2 \to 0} \frac{1}{\omega^2} \left( \sigma_k{}^A - \mathring \sigma_k{}^A \right) = - 2 \mathring H^{A B} \mathring f^C{}_i \mathring L_{BCDE} \mathring K^{DE}
\end{equation}
all other terms vanish because of the assumption of stressfreeness of the reference configuration \eqref{relaxed_background}, and $\mathring K^{DE}$ is defined by
\begin{equation}
\label{definition_K}
\mathring K^{A B} = \left. \frac{d}{d\left( \omega^2 \right)} H^{A B} \right|_{\omega^2=0} = f^A{}_i f^B{}_j \kappa^{i j}
\end{equation}
the inverse of the three-metric \eqref{three-metric} inserted into \eqref{definition_K} gives
\begin{equation}
\label{K_values}
\left( \kappa^{i j} \right) = \left(
\begin{array}{ccc}
-\frac{2 L^3}{c^2 r} & 0 & 0 \\
0 & 0 & 0 \\
0 & 0 & - \frac{1}{c^2}
\end{array}
\right)
\end{equation}
and the same components for $\mathring K^{A B}$ because of $f^A{}_i = \delta^A_i$. With this and \eqref{definition_lame}, equation \eqref{correction_definition} becomes
\begin{equation}
\label{tilde_given_by}
\left. \widetilde \sigma_i{}^A \right|_{\omega^2=0} = - \frac{1}{2} \lambda \operatorname{tr} \mathring K \delta^A_i - \mu \mathring K^{AB} \delta_{Bi}
\end{equation}
with the divergence
\begin{equation}
\mathring \nabla_A \left. \widetilde \sigma_i{}^A \right|_{\omega^2=0} = - \frac{1}{2} \lambda \left( \operatorname{tr} \mathring K \right)_{,i} - \mu \left( \mathring \nabla_A \mathring K^{AB} \right) \delta_{Bi}
\end{equation}
For further use, it is convenient to write $K^{AB}$ in cartesian coordinates, which we are free to do, since the linearized problem is defined on Euclidean space. Then \eqref{K_values} becomes
\begin{equation}
\label{K_values_cartesian}
\mathring K^{AB} = - \frac{2 L^3}{c^2} \frac{x^A x^B}{r^3} - \frac{1}{c^2} \left( \frac{\partial}{\partial \phi} \right)^A \left( \frac{\partial}{\partial \phi} \right)^B
\end{equation}
with the rotational Killing vector
\begin{equation}
\left( \frac{\partial}{\partial \varphi} \right)^A = \epsilon^{ABC} m_B X_C
\end{equation}
where $\mathbf{m}$ is the unit vector orthogonal to the rotation plane.

\subsection{Solution}
%
We now proceed as in the nonrelativistic case by shifting the coordinate origin to the center of the body at $-L \mathbf{n}$ ($\mathbf{n}$ is the unit vector pointing from the body to the gravitational centre, i.e. $(-1, 0, 0)$ in the cartesian coordinate system used above) by
\begin{equation}
\label{coordinate_shift}
\mathbf{y} = \mathbf{x} + L \mathbf{n}
\end{equation}
Thus, $\mathbf{y}$ is the normal vector to $\partial \mathcal{B} = \{ |\mathbf{y}| = a \}$, so the boundary conditions to the linearized elastostatic equation \eqref{linearized_elastic_equation} to be solved become
\begin{equation}
\label{boundary_conditions_relativistic}
\left. \left( \delta \mathring \sigma_i{}^A + \left. \widetilde \sigma_i{}^A \right|_{\omega^2=0} \right) \frac{y_A}{a} \right|_{|\mathbf{y}| = a} = 0
\end{equation}
This system is best to handle if one treats the parts of $\mathring K^{AB}$ independently by
\begin{eqnarray}
\mathring K^{AB} &=& {^g}\mathring K^{AB} + {^r}\mathring K^{AB} \\
{^g}\mathring K^{AB} &:=& - \frac{2 L^3}{c^2} \frac{x^A x^B}{r^3} \\
{^r}\mathring K^{AB} &:=& - \frac{1}{c^2} \left( \frac{\partial}{\partial \phi} \right)^A \left( \frac{\partial}{\partial \phi} \right)^B
\end{eqnarray}
Also, $\left. \widetilde \sigma_i{}^A \right|_{\omega^2=0}$ is decomposed in a similar way by inserting ${^g}\mathring K^{AB}$ and ${^r}\mathring K^{AB}$ in equation \eqref{tilde_given_by}. Thus, the elastostatic equation \eqref{linearized_elastic_equation} with the boundary condition \eqref{boundary_conditions_relativistic} decomposes into:

1. The non-relativistic equation with the homogeneous boundary conditions:
\begin{eqnarray}
\mathring \nabla_A \delta \mathring \sigma_i{}^A + f_i = 0 \\
\left. \delta \mathring \sigma_i{}^A  \frac{y_A}{a} \right|_{|\mathbf{y}| = a} = 0
\end{eqnarray}
This has already been solved

2. a relativistic correction term for $\left. \widetilde {^g}\sigma_i{}^A \right|_{\omega^2=0}$, coming from the gravitational part of the curved metric \eqref{three-metric}:
\begin{eqnarray}
\label{partial_problem_1}
\mathring \nabla_A \delta \mathring \sigma_i{}^A + \mathring \nabla_A \left. {^g} \widetilde \sigma_i{}^A \right|_{\omega^2=0} = 0 \\
\label{partial_problem_2}
\left. \left( \delta \mathring \sigma_i{}^A + \left. {^g} \widetilde \sigma_i{}^A \right|_{\omega^2=0} \right) \frac{y_A}{a} \right|_{|\mathbf{y}| = a} = 0
\end{eqnarray}
and

3. a relativistic correction term for $\left. {^r} \widetilde \sigma_i{}^A \right|_{\omega^2=0}$ of the same form as \eqref{partial_problem_1}, \eqref{partial_problem_2}, coming from the rotational part of the metric \eqref{three-metric}.

\subsection{Gravitational part}
%
We consider the relativistic corrections due to the term
\begin{equation}
{^g}K^{AB} = - \frac{2 L^3}{c^2} \frac{X^A X^B}{r^3}
\end{equation}
Its trace is given by
\begin{equation}
\operatorname{tr} {^g}K = - \frac{2 L^3}{c^2 r}
\end{equation}
so we get
\begin{equation}
{^g} \widetilde \sigma_i{}^A = \frac{\lambda L^3}{c^2} \frac{1}{r} \delta_i^A + \frac{2 \mu L^3}{c^2} \frac{x_i X^A}{r^3}
\end{equation}
The divergence of ${^g} \widetilde \sigma_i{}^A$ leads to the correction term of the force
\begin{equation}
\mathring \nabla_A \left( {^g} \widetilde \sigma_i{}^A \right) = \frac{L^3 (2 \mu - \lambda)}{c^2} \frac{x_i}{r^3} = - \mathring \nabla_i \left( \frac{L^3 (2 \mu - \lambda)}{c^2} \frac{1}{r} \right)
\end{equation}
which can be expressed as the negative gradient of the potential
\begin{equation}
U = \frac{L^3 (2 \mu - \lambda)}{c^2} \frac{1}{r} = \sum_{n=0}^\infty E_n |\mathbf{y}|^n P_n{\cos \theta}
\end{equation}
with the coefficients
\begin{equation}
\label{E_n_coefficients}
E_n = \frac{1}{c^2} (2 \mu - \lambda) L^{2-n}
\end{equation}
With these, one gets a particular solution that is of the same form as in the nonrelativistic case:
\begin{equation}
\begin{split}
\label{relativistic_partial_solution}
u_r^P &= \frac{L^3}{c^2} \frac{2 \mu - \lambda}{2 \mu + \lambda} \sum_{n=0}^\infty \frac{n+2}{2 (2n + 3)} \left( \frac{|\mathbf{y}|}{L} \right)^{n+1} P_n(\cos \theta) \\
u_\theta^P &= \frac{L^3}{c^2} \frac{2 \mu - \lambda}{2 \mu + \lambda} \sum_{n=0}^\infty \frac{1}{2 (2n + 3)} \left( \frac{|\mathbf{y}|}{L} \right)^{n+1} (-\sin \theta) P'_n(\cos \theta) \\
u_\phi^P &= 0
\end{split}
\end{equation}

Let us us now consider the boundary conditions. The unit normal vector to $\mathcal{B}$ is $y_A / a$. Inserting this into $^g \widetilde \sigma_i{}^A$ yields:
\begin{equation}
{^g} \widetilde \sigma_i := {^g} \widetilde \sigma_i{}^A \ y_A / a = \frac{\lambda L^3}{c^2} \frac{1}{r} \frac{y_i}{a} + \frac{2 \mu L^3}{c^2} \frac{x_i \left( |\mathbf{y}| - L \cos \theta \right)}{r^3} \frac{|\mathbf{y}|}{a}
\end{equation}
In order to plug this in our ansatz, we need to develop this in suitable (associated) Legendre functions of the form
\begin{eqnarray}
\label{gsigmar}
\left. {^g} \widetilde \sigma_r \right|_{|\mathbf{y}|=a} &=& \sum_{n=0}^\infty M_n \left( \frac{a}{L} \right)^n P_n( \cos \theta ) \\
\label{gsigmatheta}
\left. {^g} \widetilde \sigma_\theta \right|_{|\mathbf{y}|=a} &=& \sum_{n=0}^\infty N_n \left( \frac{a}{L} \right)^n P'_n( \cos \theta ) (- \sin \theta)
\end{eqnarray}
The most convenient way to accomplish this is to start by the generating function
\begin{equation}
\frac{1}{r} = \frac{1}{\sqrt{|\mathbf{y}|^2 - 2 L |\mathbf{y}| \cos \theta + L^2}} = \frac{1}{L} \sum_{n=0}^\infty \left( \frac{|\mathbf{y}|}{L} \right)^n P_n(\cos \theta)
\end{equation}
and differentiate it with respect to $|\mathbf{y}|$, which yields
\begin{equation}
\frac{L \cos \theta - |\mathbf{y}|}{\sqrt{|\mathbf{y}|^2 - 2 L |\mathbf{y}| \cos \theta + L^2}^3} = \frac{1}{L^2} \sum_{n=1}^\infty n \left( \frac{|\mathbf{y}|}{L} \right)^{n-1} P_n(\cos \theta)
\end{equation}
The components of ${^g} \widetilde \sigma_i$ in spherical coordinates then become (again, we rename $|\mathbf{y}|$ to $r$ from now on to save some writing effort)
\begin{equation}
\begin{split}
\left. {^g} \widetilde \sigma_r \right|_{|\mathbf{y}|=a} &:= \left. {^g} \widetilde \sigma_i e_r^i \right|_{|\mathbf{y}|=a} = \frac{\lambda L^2}{c^2} \sum_{n=0}^\infty \left( \frac{a}{L} \right)^n P_n(\cos \theta) - \\
& \qquad - \frac{2 \mu L^2}{c^2} \left( \frac{a}{L} - \cos \theta \right) \sum_{n=1}^\infty n \left( \frac{a}{L} \right)^{n-1} P_n(\cos \theta)  \\
\left. {^g} \widetilde \sigma_\theta \right|_{|\mathbf{y}|=a} &:= \left. {^g} \widetilde \sigma_i e_\theta^i \right|_{|\mathbf{y}|=a} = - \frac{2 \mu L^2}{c^2} \sin \theta \sum_{n=1}^\infty n \left( \frac{a}{L} \right)^{n-1} P_n(\cos \theta)  \\
\left. {^g} \widetilde \sigma_\phi \right|_{|\mathbf{y}|=a} &:= {^g} \widetilde \sigma_i e_\phi^i = 0
\end{split}
\end{equation}
To get ${^g} \widetilde \sigma_r$ in the desired form \eqref{gsigmar}, we use the identity (see \cite{whittaker-watson})
\begin{equation}
x P_n(x) = \frac{1}{2 n + 1} \left( (n+1) P_{n+1} + n P_{n-1} \right)
\end{equation}
which holds for all $n \ge 1$, to get
\begin{equation}
\begin{split}
\left. {^g} \widetilde \sigma_r \right|_{|\mathbf{y}|=a} &= \frac{2 \mu L^2}{c^2} \left( \sum_{n=0}^\infty \left( \frac{\lambda}{2 \mu} - n \right) \left( \frac{a}{L} \right)^n P_n(\cos \theta) \right. \\
& \qquad \left. + \sum_{n=1}^\infty \left( \frac{a}{L} \right)^{n-1} \left( \frac{n (n+1)}{2 n + 1} P_{n+1}(\cos \theta) + \frac{n^2}{2 n + 1} P_{n-1}(\cos \theta) \right) \right)
\end{split}
\end{equation}
After two index-shifts, it becomes into the form \eqref{gsigmar} with the coefficients
\begin{equation}
\label{M_n}
M_n = \frac{2 \mu L^2}{c^2} \left( \frac{\lambda}{2 \mu} + \left( \frac{L}{a} \right)^2 \frac{n (n-1)}{2 n - 1} + \frac{(n+1)^2}{2 n + 3} - n \right)
\end{equation}
In an analogue way, for ${^g} \widetilde \sigma_\theta$, we use the identity (also for all $n \ge 1$ and taken from \cite{whittaker-watson} as well)
\begin{equation}
P_n(x) = \frac{1}{2 n + 1} \left( P_{n+1}' - P_{n-1}' \right)
\end{equation}
to get (after two index-shifts as well) it into the form \eqref{gsigmatheta} with
\begin{equation}
\label{N_n}
N_n = \frac{2 \mu L^2}{c^2} \left( \left( \frac{L}{a} \right)^2 \frac{n - 1}{2 n - 1} - \frac{n + 1}{2 n + 3} \right)
\end{equation}

We now have everything ready to plug into the ansatz from Papkowitsch and Neuber, similar to the nonrelativistic case. The particular solution is given by the same formula as in the nonrelativistic case, but with the coefficients \eqref{E_n_coefficients}. Also the homogeneous solution is the same as in the nonrelativistic case. The free constants $A_n$, $B_n$ and $C_n$ have to be determined from the boundary conditions
\begin{align}
0 &= \left. \left( \sigma^H_{r r} + \sigma^P_{r r} + {^g} \widetilde \sigma_r \right) \right|_{r=a}  \\
0 &= \left. \left( \sigma^H_{r \theta} + \sigma^P_{r \theta} + {^g} \widetilde \sigma_\theta \right) \right|_{r=a}  \\
0 &= \left. \left( \sigma^H_{r \phi}     \right)           \right|_{r=a}
\end{align}
From the boundary condition for $\sigma^H_{r \phi}$ and the discrete symmetries imposed, we conclude that $u_\phi=0$, as in the nonrelativistic case. The required stress tensor components for the particular solution are
\begin{align}
\sigma^P_{rr} &= 2 \mu \frac{L^2}{c^2} \frac{2 \mu - \lambda}{2 \mu + \lambda} \sum_{n=0}^\infty \left( \frac{(n+1) (n+2)}{2 (2n+3)} + \frac{\lambda}{2 \mu} \right) \left( \frac{r}{L} \right)^n P_n (\cos \theta) \\
\sigma^P_{r \theta} &= 2 \mu \frac{L^2}{c^2} \frac{2 \mu - \lambda}{2 \mu + \lambda} \sum_{n=0}^\infty \frac{n+1}{2 (2n+3)} \left( \frac{r}{L} \right)^n (- \sin \theta) P_n'(\cos \theta)
\end{align}
With this and \eqref{gsigmar}, \eqref{gsigmatheta}, \eqref{M_n}, \eqref{N_n}, the other two boundary conditions become
\begin{multline}
\label{linear_system_for_A_B_relativistic_r}
(n + 1) \left( n^2 - n - \frac{3 \lambda + 2 \mu}{\lambda + \mu} \right) a^n A_n + n (n - 1) a^{n-2} B_n = \\
= \frac{L^2}{c^2} \left( \frac{a}{L} \right)^n \left( \frac{3 \lambda + 2 \mu}{\lambda + 2 \mu} \quad \frac{n^2-n-4}{2 (2 n + 3)} - \left( \frac{L}{a} \right)^2 \frac{n (n-1)}{2n-1}   \right)
\end{multline}
\begin{multline}
\label{linear_system_for_A_B_relativistic_theta}
\left( n^2 + 2 n - \frac{\mu}{\lambda + \mu} \right) a^n A_n + (n - 1) a^{n-2} B_n =  \\
= \frac{L^2}{c^2} \left( \frac{a}{L} \right)^n \left( \frac{3 \lambda + 2 \mu}{\lambda + 2 \mu} \quad \frac{n+1}{2 (2 n + 3)} - \left( \frac{L}{a} \right)^2 \frac{(n-1)}{2n-1}  \right)
\end{multline}
For $n=0$, we actually have just one equation, because the $\sigma_{r \theta}$ component of both the particular and the homogeneous solution vanish identically, and $B_0$ does not occur in the overall solution. we thus get
\begin{equation}
A_0 = \frac{2}{3} \frac{L^2}{c^2} \frac{\lambda + \mu}{\lambda + 2 \mu}
\end{equation}
which leads to
\begin{align}
u_r^0 &= - \frac{1}{3} \frac{L^2}{c^2} r \\
u_\theta^0 &= 0
\end{align}
This means that we get a contraction of the elastic sphere due to relativistic effects.

To get a solution of the system \eqref{linear_system_for_A_B_relativistic_r}, \eqref{linear_system_for_A_B_relativistic_theta} for $n>0$, we view the determinant of the system. Like in the non-relativistic case, it is
\begin{equation}
D_n = - (n - 1) \left( 2 n (n - 1) + \frac{3 \lambda + 2 \mu}{\lambda + \mu} (2 n + 1) \right) a^{2n-2}
\end{equation}
$D_1=0$, so there is no unique solution for $n=1$, but we see that equation \eqref{linear_system_for_A_B_relativistic_r} becomes $-2$ times equation \eqref{linear_system_for_A_B_relativistic_theta}. Thus there are solutions, which are given by
\begin{equation}
A_1 = \frac{1}{5} \frac{L}{c^2} \frac{\lambda + \mu}{\lambda + 2 \mu}
\end{equation}
while $B_1$ is arbitrary. After adding the $n=1$ term of the particular solution \eqref{relativistic_partial_solution} to the homogeneous solution, we thus get
\begin{align}
u_r^1 &= - \frac{1}{10} \frac{L}{c^2} r^2 \cos \theta + B_1 \cos \theta \\
u_\theta^1 &= \frac{7}{10} \frac{L}{c^2} r^2 (-\sin \theta) + B_1 (- \sin \theta)
\end{align}
The term proportional to $B_1$ is the translational Killing vector $e_z^i$, which has to be determined from the equilibrium conditions after the rest of the solution, just as in the nonrelativistic case.

For $n>1$, we have $D_n<0$, so we can solve the system \eqref{linear_system_for_A_B_relativistic_r}, \eqref{linear_system_for_A_B_relativistic_theta} directly to get
\begin{align}
\label{relativistic_constants_for_homogeneous_solution}
A_n &= \frac{L^3}{c^2} \frac{1}{L^{n+1}} \frac{(n+2) \frac{3 \lambda + 2 \mu}{\lambda + 2 \mu}}{(2 n + 3) \left( 2 n (n-1) + \frac{3 \lambda + 2 \mu}{\lambda + \mu} (2 n + 1) \right)} \\
B_n &= - \frac{L^3}{c^2} \frac{1}{L^{n-1}} \left( \frac{1}{2 n-1} + \left( \frac{a}{L} \right)^2 \frac{\frac{3 \lambda + 2 \mu}{\lambda + \mu}}{2 \left( 2 n (n-1) + \frac{3 \lambda + 2 \mu}{\lambda + \mu} (2 n + 1) \right)} \right)
\end{align}
After adding the particular solution \eqref{particular_solution} and the homogeneous solution \eqref{homogeneous_solution} using the constants \eqref{relativistic_constants_for_homogeneous_solution} and \eqref{E_n_coefficients}, we thus get as overall solution
\begin{equation}
\begin{split}
\label{whole_solution_relativistic}
u_r &= s \sum_{n=2}^\infty \left( F_n \left( \frac{r}{L} \right)^{n+1} + G_n \frac{a^2}{L^2} \left( \frac{r}{L} \right)^{n-1} \right) P_n(\cos \theta) \\
u_\theta &= s \sum_{n=2}^\infty \left( H_n \left( \frac{r}{L} \right)^{n+1} + I_n \frac{a^2}{L^2} \left( \frac{r}{L} \right)^{n-1} \right) \frac {d P_n}{d \theta}(\cos \theta) \\
u_\phi &= 0
\end{split}
\end{equation}
with
\begin{equation}
\begin{split}
s   &= \frac{\omega^2 L^3}{c^2} \\
F_n &= \frac{ (n+2) \left( 2 n^2 + \frac{\lambda}{\lambda + \mu} n - \frac{3 \lambda + 2 \mu}{2 (\lambda + \mu)} \right)}{\left( 2n + 3 \right) \left( 2 n (n - 1) + \frac{3 \lambda + 2 \mu}{\lambda + \mu} (2n + 1) \right)} \\
G_n &= n I_n \\
H_n &= \frac{2 n^2 + \frac{13 \lambda + 8 \mu}{\lambda + \mu} n + \frac{11 (3 \lambda + 2 \mu)}{2 (\lambda + \mu)}}{(2 n + 3) \left( 2 n (n - 1) + \frac{3 \lambda + 2 \mu}{\lambda + \mu} (2n + 1) \right)} \\
I_n &= - \left( \frac{L}{a} \right)^2 \frac{1}{2 n-1} - \frac{\frac{3 \lambda + 2 \mu}{\lambda + \mu}}{2 \left( 2 n (n-1) + \frac{3 \lambda + 2 \mu}{\lambda + \mu} (2 n + 1) \right)}
\end{split}
\end{equation}
%

%
\subsection{Rotational part}
%
We consider the relativistic corrections due to the term
\begin{equation}
\label{rKAB}
{^r} \mathring K^{AB} = - \frac{1}{c^2} ( \mathbf{m} \times \mathbf{x} )^A ( \mathbf{m} \times \mathbf{x} )^B
\end{equation}
Its trace is given by
\begin{equation}
\label{trace_K}
\operatorname{tr} {^r} \mathring K = - \frac{1}{c^2} \left( \mathbf{x}^2 - (\mathbf{x}, \mathbf{m})^2 \right)
\end{equation}
For the sake of completeness, we collect the terms \eqref{tilde_given_by}, \eqref{rKAB} and \eqref{trace_K} to get
\begin{equation}
\left. {^r} \widetilde \sigma_i{}^A \right|_{\omega^2=0} = \frac{1}{c^2} \left( \lambda \left( \frac{1}{2} \left( \mathbf{x}^2 - (\mathbf{x}, \mathbf{m})^2 \right) \right) \delta_i^A + \mu \left( \left( \frac{\partial}{\partial \phi} \right)^A \left( \frac{\partial}{\partial \phi} \right)^B \right) \delta_{B i} \right)
\end{equation}
To calculate the force correction term in the elastostatic equation \eqref{linearized_elastic_equation}, we calculate
\begin{equation}
- \frac{1}{2} \lambda \left( \operatorname{tr} {^r} \mathring K \right)_{,i} = \lambda \frac{1}{c^2} \left( x_i - (\mathbf{x}, \mathbf{m}) m_i \right)
\end{equation}
and the divergence of ${^r} \mathring K^{AB}$ is given by
\begin{equation}
\mathring \nabla_B {^r} \mathring K^{AB} = \frac{1}{c^2} \left( x^A - (\mathbf{m}, \mathbf{x} ) m^A \right)
\end{equation}
Thus the correction term in \eqref{linearized_elastic_equation} becomes
\begin{equation}
\mathring \nabla_A \left. {^r} \widetilde \sigma_i{}^A \right|_{\omega^2=0} = \frac{\lambda - \mu}{c^2} \left( x_i - (\mathbf{x}, \mathbf{m}) m_i \right)
\end{equation}
After the coordinate origin shift \eqref{coordinate_shift}, we get
\begin{equation}
{^r} \mathring K^{AB} = - \frac{1}{c^2} ( \mathbf{m} \times (\mathbf{y} - L \mathbf{n}) )^A ( \mathbf{m} \times (\mathbf{y} - L \mathbf{n}) )^B
\end{equation}
This can be even further decomposed into
\begin{align}
{^{r1}} \mathring K^{AB} &= - \frac{1}{c^2} ( \mathbf{m} \times \mathbf{y})^A ( \mathbf{m} \times \mathbf{y} )^B \\
{^{r2}} \mathring K^{AB} &= - \frac{L}{c^2} \left( -( \mathbf{m} \times \mathbf{n} )^A ( \mathbf{m} \times \mathbf{y})^B -( \mathbf{m} \times \mathbf{y} )^A ( \mathbf{m} \times \mathbf{n})^B \right. \\
& \qquad \left. + L ( \mathbf{m} \times \mathbf{n} )^A ( \mathbf{m} \times \mathbf{n})^B \right)
\end{align}
The ${^{r1}} \mathring K^{AB}$ part leads to (in a cartesian coordinate system with $\mathbf{m}$ as the $z$-axis) the system
\begin{equation}
\label{rigid_rotation}
\begin{split}
\mu \Delta \delta \phi^i + \left( \lambda + \mu \right) \operatorname{grad} \operatorname{div} \delta \phi^i + \left( \rho + \frac{\lambda - \mu}{c^2} \right) \left( \begin{array}{c} y^1 \\ y^2 \\ 0 \end{array} \right)^i = 0 \\
\left. \left( \delta \mathring \sigma_i{}^A y_A + \frac{\lambda}{2 c^2} \left( \mathbf{y}^2 - (\mathbf{y}, \mathbf{m})^2 \right) y_i \right) \right|_{|\mathbf{y}| = a} = 0
\end{split}
\end{equation}
This describes spheres in rigid rotations along an axis through their center; a solution is given in \cite{beig-schmidt-1}.

It remains to find a solution to ${^{r2}} \mathring K^{AB}$. By looking at it the right way, it becomes immediately clear that it can be considered as the strain tensor belonging to the followin perturbation:
\begin{equation}
\label{delta_f_trivial}
\delta f^A = - \frac{L}{c^2} \left( \frac{L}{2} \left( \mathbf{m} \times \mathbf{n} \right)^A \left( \mathbf{m} \times \mathbf{n} \right)^j y_j - (\mathbf{m} \times \mathbf{n})^j y_j (\mathbf{m} \times \mathbf{y} )^A \right)
\end{equation}
i.e.
\begin{equation}
{^{r2}}K^{AB} = \delta f^A{}_i f^B{}_j \delta^{ij} + f^A{}_i \delta f^B{}_j \delta^{ij}
\end{equation}
as can be easily seen by elementary vector calculus (Remember that the flat metric $\delta^{ij}$ is used instead of the curved one $h^{ij}$, because after linearization, the whole problem is considered to be defined on Euclidean space). Thus, the displacement belonging to minus the one given in equation \eqref{delta_f_trivial} has a strain tensor that cancels out ${^{r2}} \mathring K^{AB}$, and thus automatically satisfies both the elastostatic equation and the boundary conditions belonging to ${^{r2}} \widetilde \sigma_i{}^A$. So the displacement vector is basically identical to \eqref{delta_f_trivial} (going from $\delta f^A$ to $\delta \phi^i$ causes another minus sign) apart from the factor $\omega^2$:
\begin{equation}
u^i = - \frac{L \omega^2}{c^2} \left( \frac{L}{2} \left( \mathbf{m} \times \mathbf{n} \right)^i \left( \mathbf{m} \times \mathbf{n} \right)^j y_j - (\mathbf{m} \times \mathbf{n})^j y_j (\mathbf{m} \times \mathbf{y} )^i \right)
\end{equation}
In a cartesian coordinate system with the axes $(\mathbf{m}, \mathbf{n} \times \mathbf{m}, \mathbf{n})$, this means:
\begin{equation}
(u^i) = \frac{\omega^2 L^2}{c^2} y \left( \begin{array}{c}
0 \\
\frac{z}{L} - \frac{1}{2} \\
-\frac{y}{L}
\end{array} \right)
\end{equation}
Note that this term breaks the rotational symmetry.

\begin{figure}[hp]
\includegraphics{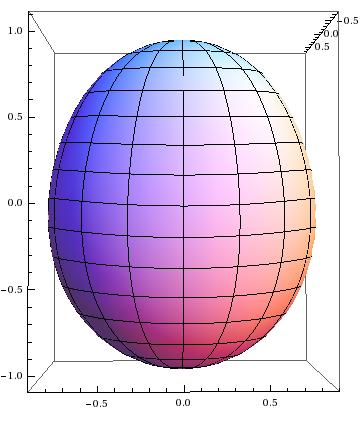}
\caption{A graphic display of the solution including all relativistic correction terms, viewed in direction $\mathbf{m}$. The material and orbit parameters are the same as for the nonrelativistic picture \ref{nichtrelativistische_graphik} on page \pageref{nichtrelativistische_graphik}: The material used is steel ($\lambda$ = 108 GPa, $\mu$ = 78 GPa, $\rho_0$ = 7860 kg/$m^3$). For the orbit velocity $\omega L$ = 9.3 km/s was used, and $L=3a$. The gravitational centre is in the direction of the top side of the picture. In order to see the relativistic effects well, the speed of light was set to $c$=20.8 km/s.}
\label{relativistische_graphik_1}
\end{figure}

\clearpage

\begin{figure}[hp]
\includegraphics{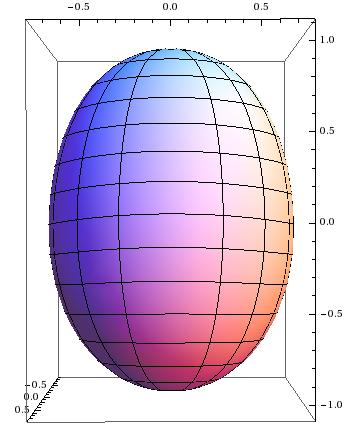}
\caption{A graphic display of the solution \eqref{whole_solution}, viewed in direction $\mathbf{m} \times \mathbf{n}$. Otherwise, the picture is similar to figure \ref{relativistische_graphik_1}.}
\label{relativistische_graphik_2}
\end{figure}

\clearpage

%
\section{Acknowledgement}
%
Many thanks are due to Robert Beig for countless ideas and fruitful discussions. Helpful remarks by Mark Heinzle are also gratefully acknowledged.
%
%

\end{document}